 \documentclass[lettersize,journal, onecolumn, draftclsnofoot]{IEEEtran}

\ifCLASSINFOpdf
\else
\fi
\usepackage[numbers]{natbib} 
\usepackage{xcolor}
\usepackage{amsmath} 
\usepackage{graphicx}
\usepackage{tabulary}
\usepackage{tabularx}
\usepackage{makecell}
\usepackage{booktabs}
\usepackage{multirow}
\usepackage{multicol}
\usepackage[ruled, vlined]{algorithm2e}
\usepackage{hyperref}
\usepackage{float}
\usepackage{nomencl}

\usepackage{rotating}
\usepackage{tikz}

\definecolor{green(munsell)}{rgb}{0.0, 0.66, 0.47}

\begin{document} 
\title{Valuation of Public Bus Electrification with Open Data} 

\author{
\IEEEauthorblockN{Upadhi Vijay\IEEEauthorrefmark{1},
Soomin Woo\IEEEauthorrefmark{1}\IEEEauthorrefmark{5}, Scott J. Moura\IEEEauthorrefmark{1},
Akshat Jain\IEEEauthorrefmark{1}, David Rodriguez\IEEEauthorrefmark{2}, Sergio Gambacorta\IEEEauthorrefmark{3}, Giuseppe Ferrara\IEEEauthorrefmark{3}, Luigi Lanuzza\IEEEauthorrefmark{3}, Christian Zulberti\IEEEauthorrefmark{4}, Erika Mellekas\IEEEauthorrefmark{2}, and Carlo Papa\IEEEauthorrefmark{4}
}\\ 
\IEEEauthorblockA{\IEEEauthorrefmark{1} 
University of California, Berkeley, USA, \IEEEauthorrefmark{2}Enel X, North America, Inc., USA, \IEEEauthorrefmark{3}Enel X, Italy, and \IEEEauthorrefmark{4}Enel Foundation, Italy\\
\IEEEauthorrefmark{5}Corresponding Author 
}
\thanks{Email: U.Vijay (upadhi\_vijay@berkeley.edu), S.Woo (soomin.woo@berkeley.edu), S.J.Moura (smoura@berkeley.edu), A.Jain (akshatj@berkeley.edu), D.Rodriguez (david.rodriguez2@enel.com), S.Gambacorta (sergio.gambacorta@enel.com), G.Ferrara (giuseppe.ferrara4@enel.com), L.Lanuzza (luigi.lanuzza@enel.com), C. Zulberti (christian.zulberti@enel.com), E.Mellekas (erika.mellekas@enel.com), C.Papa (carlo.papa@enel.com)}
\thanks{Detailed Affiliation: U.Vijay, S.Woo, and S.J.Moura are at Department of Civil and Environmental Engineering, University of California-Berkeley, Davis Hall, Berkeley, California, 94720, USA. A.Jain is at Department of Electrical Engineering and Computer Sciences, University of California-Berkeley, Soda Hall, Berkeley, California, 94720, USA. D.Rodriguez and E.Mellekas are at Enel X, North America, Inc., One Marina Park Drive, Boston, 02210, MA, USA. S. Gambacorta is at Enel X, Innovation and Sustainability Global, Smart City, Viale Tor di Quinto, Rome, 00191, Italy. G.Ferrara is at Enel X, Innovation and Sustainability Global, Smart City, Passo Martino, Catania, 95121, Italy. L.Lanuzza is at Enel X, Innovation and Sustainability B2C \& B2B Innovation Factory, Viale Tor di Quinto, Rome, 00191, Italy. C.Zulberti and C.Papa are at Enel Foundation, Via Bellini, Rome, 00198, Italy.}
} 
\maketitle     
\begin{abstract}
This research provides a novel framework to estimate the economic, environmental, and social values of electrifying public transit buses, for cities across the world, based on open-source data. Electric buses are a compelling candidate to replace diesel buses for the environmental and social benefits. However, the state-of-art models to evaluate the value of bus electrification are limited in applicability because they require granular and bespoke data on bus operation that can be difficult to procure. Our valuation tool uses General Transit Feed Specification, a standard data format used by transit agencies worldwide, to provide high-level guidance on developing a prioritization strategy for electrifying a bus fleet. We develop physics-informed machine learning models to evaluate the energy consumption, the carbon emissions, the health impacts, and the total cost of ownership for each transit route. We demonstrate the scalability of our tool with a case study of the bus lines in the Greater Boston and Milan metropolitan areas.
\end{abstract}

\IEEEpeerreviewmaketitle

\makenomenclature

\nomenclature{$\text{TCO}_{d,r}^\text{NPV}$}{Net present value of Total Cost of Ownership of the diesel bus fleet for route $r$ in USD}
\nomenclature{$\text{TCO}_{e,r}^\text{NPV}$}{Net present value of Total Cost of Ownership of the electric bus fleet for route $r$ in USD}

\nomenclature{$N_{r, b}(s,t)$}{Estimated number of buses at time $t$ and stop $s$ on the route $r$ for direction $b$}
\nomenclature{$N_{r, b}^v$}{Numbers of buses estimated on route $r$ for direction $b$}

\nomenclature{$C_{b}(s,t)$}{Cycle length estimated for the bus that arrives at stop $s$ at time $t$ for the direction $\text{b}$ in minutes}
 
\nomenclature{$h_{b}(s,t)$}{Headway estimated for the bus that arrives at stop $s$ at time $t$ for the direction $\text{b}$ in minutes} 

\nomenclature{$N^v_r$}{Number of buses needed for route $r$}
\nomenclature{$\overline{v_{\mathcal{C}_{k,r}}}$}{Average velocity of bus for $k$-th cluster of route $r$, $\mathcal{C}_{k,r}$, in m/s}
\nomenclature{$\overline{v_r}$}{Average velocity of bus for route $r$ in m/s}
\nomenclature{$\alpha^{i,j}$}{Road grade between two points, $i$ and $j$, in degrees}
\nomenclature{$\Delta e_{i,j}$}{Change in elevation between two points, $i$ and $j$ in km}
\nomenclature{$d_{i,j}$}{Distance from point $i$ to point $j$ in km}

\nomenclature{$\mathcal{C}_{k,r}$}{$k$-th cluster of route $r$} 

\nomenclature{$\Delta t$}{Time interval in the Manhattan drive cycle, $0.1 sec$}

\nomenclature{$N^c_r$}{Number of chargers needed for route $r$}
\nomenclature{$\overline{E_{\mathcal{C}_{k,r}}}$}{Daily energy consumption for $k$-th cluster of route $r$, $\mathcal{C}_{k,r}$, in kWh}
\nomenclature{$E_j$}{Energy consumption of a single trip $j$ in kWh}
\nomenclature{$\Delta E_i$}{Energy consumption to traverse from point $i$ to the consecutive point $j$ over a time step $\Delta t$ in kWh}

\nomenclature{$P_c$}{Nominal charging power, 50kW}
\nomenclature{$T_c$}{Fastest charging time in hours, 4.5 to 5 for a 40-foot bus \cite{bydbus}}
\nomenclature{$\eta_c$}{Charging efficiency (user input)}
\nomenclature{$E_{max,b}$}{Energy capacity of battery, $352$kWh for a 40-foot bus \cite{bydbus}}

\nomenclature{$\text{CAPEX}_{e,r}$}{Capital cost of electric buses for route $r$ in USD}
\nomenclature{$\text{CAPEX}_{d,r}$}{Capital cost of diesel buses for route $r$ in USD}
\nomenclature{$c_{e,bus}$}{Unit cost of an electric bus in USD (user input)}
\nomenclature{$c_{c,install}$}{Installation cost of a charger, 17,692 USD per charger \cite{nicholas2019estimating}}
\nomenclature{$c_{c,unit}$}{Unit cost of a charger, 27,549 USD per charger \cite{nicholas2019estimating}}
\nomenclature{$c_{d,bus}$}{Unit cost of a diesel bus, for a 40-foot bus: 485,000 USD in the US 360,000 USD in Italy \cite{tong2017life}}

\nomenclature{$\text{OPEX}_e^y$}{Operation cost of electric buses for year $y$ in USD}
\nomenclature{$C_{kWh,r}^y$}{Annual energy cost of route $r$ for year $y$ in USD}
\nomenclature{$C_{DC,r}^y$}{Annual demand charge of route $r$ for year $y$ in USD}

\nomenclature{$C_{O\&M,e,r}^y$}{Annual operation and maintenance cost of electric buses for route $r$ for year $y$ in USD}
\nomenclature{$c_{kWh}$}{Electricity price per energy in USD per kWh (user input)}
\nomenclature{$E_r^\text{annual}$}{Annual energy consumption of route $r$ in kWh}
\nomenclature{$r_{kWh}$}{Percentage change in energy price each year, -0.1\% for Boston \cite{Johnson2020} and 1.1\% for Milan \cite{europeancommission}}
\nomenclature{$c_{DC}$}{Demand charge in USD per kW (user input)}

\nomenclature{$r_{DC}$}{Percentage change in demand charge each year (user input)}
\nomenclature{$c_{O\&M,e}^v$}{Operation and maintenance cost of an electric bus, 0.64 USD per mile \cite{Johnson2020}}
\nomenclature{$\text{VKT}_r^\text{annual}$}{Annual vehicle kilometers traveled for route $r$}
\nomenclature{$c_{O\&M}^c$}{Operation and maintenance cost of a charger, 500 USD per charger per year \cite{blynn2018accelerating}}

\nomenclature{$\text{OPEX}_d^y$}{Operation cost of diesel buses for year $y$}
\nomenclature{$C_\text{fuel,r}^y$}{Annual fuel cost of route $r$ for year $y$ in USD}
\nomenclature{$C_{O\&M,d,r}^y$}{Annual operation and maintenance cost of diesel buses for route $r$ for year $y$ in USD}
\nomenclature{$c_{fuel}$}{Fuel price in USD per km (user input)} 
\nomenclature{$r_f$}{Percentage change in fuel price each year, 0.7\% for Boston and 4.3\% for Milan \cite{Johnson2020}}
\nomenclature{$c_{O\&M,d}^v$}{Operation and maintenance cost of a diesel bus, 0.88 USD per mile \cite{Johnson2020}}

\nomenclature{$C_{d,salv,r}$}{Salvage cost of diesel buses for route $r$ in USD}
\nomenclature{$C_{e,salv,r}$}{Salvage cost of electric buses for route $r$ in USD}
\nomenclature{$r_v$}{Residual value of a bus, 15\% \cite{hensher2007bus}}
\nomenclature{$r_{c}$}{Residual value of a charger, 15\% \cite{hensher2007bus}}
 
\nomenclature{$r_d$}{Discount rate, 3.5\%}

\nomenclature{$P_{trac}^{i,j}$}{Tractive power from point $i$ to point $j$ in Watt}
\nomenclature{$v_i$}{Velocity of the bus estimated at point $i$ in m/s}
  
\nomenclature{$\overline{C_{\mathcal{C}_{k,r}}}$}{Sample mean of the cycle lengths for trips in cluster $\mathcal{C}_{k,r}$}

\nomenclature{$F_{accel}^{i,j}$}{Acceleration force from point $i$ to point $j$ in Newtons}
\nomenclature{$F_{grade}^{i,j} $}{Grade climbing force from point $i$ to point $j$ in Newtons}
\nomenclature{$F_{roll}^{i,j}$}{Rolling resistance force from point $i$ to point $j$ in Newtons}
\nomenclature{$F_{drag}^{i,j}$}{Air drag force from point $i$ to point $j$ in Newtons}

\nomenclature{$m$}{Total mass of the bus in kg}

\nomenclature{$a_i$}{Acceleration estimated at point $i$ in $m/{s^2}$}
\nomenclature{$g$}{Gravitational acceleration, $9.81m/{s^2}$} 
\nomenclature{$C_{rr}$}{Coefficient of rolling resistance, $0.00697$ \cite{franca2018electricity}}
\nomenclature{$\rho$}{Air density, $1.2kg/m^3$ \cite{ElTaweel2021}}
\nomenclature{$A$}{cross-sectional area of the bus, $8.78m^2$ for a 40-foot bus \cite{bydbus}}
\nomenclature{$C_d$}{Air drag coefficient, $0.65$ \cite{rasu2016cfd}} 
\nomenclature{$m_{bus}$}{Mass of the bus, 14,050 kg for a 40-foot  bus \cite{bydbus}}
\nomenclature{$m_{pass}$}{Average passenger mass, $70kg$ per person \cite{Pan2019}}
\nomenclature{$\beta_{t}^{pass}$}{Number of passengers at time $t$}
\nomenclature{$\eta_{motor}$}{Motor efficiency, $0.85$ \cite{Pan2019}} 
\nomenclature{$\eta_{battery}$}{Battery efficiency, $0.95$ \cite{Pan2019}} 
\nomenclature{$Q_t^{HVAC}$}{Power consumption for the HVAC system at time $t$ \cite{ElTaweel2021}}
\nomenclature{$\eta^{COP}$}{HVAC system coefficient of performance, $2$ \cite{ElTaweel2021}} 

\nomenclature{$P_{aux}$}{Auxiliary power, 2kW \cite{Vepslinen2019}} 
\nomenclature{$P_{non-trac}^t$}{Non-tractive power at time $t$ in Watt}

\nomenclature{$P_{max,motor}$}{Electric motor power capacity, $300$kW for a 40-foot bus \cite{bydbus}}

\nomenclature{$EE_s$}{Energy efficiency between a pair of consecutive stops $s$ in kWh per km} 

\nomenclature{$c_v$}{Distance unit conversion factor, $0.621371$ miles $\cdot \text{km}^{-1}$}
 
\nomenclature{$FE$}{Fuel economy of a diesel bus in miles per gallon}
\nomenclature{$\text{CO}_2^{d,r}$}{Well-to-wheel $\text{CO}_2$ emissions of a diesel bus for a given route $r$ in kg}

\nomenclature{$\text{CO}_2^{e,r}$}{Well-to-wheel $\text{CO}_2$ emissions of an electric bus for a given route $r$ in kg}

\nomenclature{$EF^{CO_2, d}_{W2T}$}{Well-to-tank emission factor of $\text{CO}_2$ for a diesel bus,  $310$g/km for Boston \cite{FTA2016} and  $149.1$g/km for Milan \cite{ribau2014efficiency}}

\nomenclature{$EF^{CO_2, e}_{W2T}$}{Well-to-tank emission factor of $\text{CO}_2$ for an electric bus,  0.2369 kg/kWh for Boston \cite{EPA2018} and  0.483 kg/kWh for Milan \cite{saheb2014develop}}

\nomenclature{$EF^{CO_2, d}_{T2W}$}{Tank-to-wheel emission factor of $\text{CO}_2$ for a diesel bus, $10.21$kg/gallon \cite{jamriska2004diesel} }

\nomenclature{$PM^{d,r}$}{PM2.5 exhaust for diesel buses for a given route $r$ in grams}
\nomenclature{$EF^{PM2.5, d}_{T2W}$}{Tank-to-wheel emission factor of PM2.5 for a diesel bus, $0.583$g/km \cite{EPA2020}}
\nomenclature{$HI_r$}{Health impact of the PM2.5 emissions of a diesel bus for a given route $r$ in USD}

\nomenclature{$PI$}{Population intake in kg of PM2.5 inhaled by human}
\nomenclature{$FF$}{Effect factor in disability-adjusted life years lost per kg of PM2.5 inhaled,  $260.110$ for Boston and  $79.802$ for Milan\cite{Fantke2019}}
\nomenclature{$VSL$}{Value of statistical life in  million US dollars,  $6.267$ for Boston and  $4.303$ for Milan converted to the net present values of 2020 from \cite{miller2000variations}}
\nomenclature{$IF$}{Intake fraction in ppm,  $25.8$ for Boston and $35.3$  for Milan \cite{Apte2012}}
 
\printnomenclature

\section{Introduction}
\label{sec:intro}
This paper proposes a novel model to estimate the economic, environmental, and social values of electrifying public transit buses for cities worldwide by using a global standard of open transit data. Vehicle electrification is crucial for reducing the climate impact of the transportation sector, which currently accounts for 16.2\% of the global greenhouse gas emissions \cite{EIA}. Zero-emission electric vehicles can significantly improve the air quality, health, and environmental equity \cite{Tessum2019}, \cite{CARB}. In particular, electric buses have shown superior performance in energy efficiency and emissions than diesel buses \cite{Jwa2018}.

However, the state-of-art valuation models are challenging to use as they require detailed data on transit operation, such as the bus route and schedule, the passenger demand, the driving speed profile, the road elevation, and the charging infrastructure information. The collection of such data can be arduous and sometimes impractical. Based on our literature review, no study has developed a generalized tool that evaluates the benefits of bus electrification across multiple cities and routes via open data so as to determine the target city for further analysis. With an innovative modeling methodology that uses open-source data, we provide a valuation tool to aid the decision-making process of strategically identifying bus routes to electrify across transit agencies worldwide.

\subsection{Literature Review}
To the best of our knowledge, our research is unique in using only open-source data for bus electrification valuation. Nevertheless, there is a rich body of related research, described next and organized into five categories.

\subsubsection{Energy consumption modeling of electric buses}
Various studies model the energy consumption of electric vehicles using a large number of physical parameters. In an experiment with real-world driving conditions, Younes et al. found that the ambient temperature, driving style, and route type strongly correlated with battery energy consumption \cite{Younes2013}. Vepslinen et al. found that the energy consumption of electric city buses depended on the variations and uncertainty in transit operation, such as driving style, route, traffic, ambient temperature, rolling resistance, drive efficiency, auxiliary power, vehicle mass, and elevation profile \cite{Vepslinen2018}. 

Many studies estimate energy consumption with bespoke data on the specific routes and bus powertrains. For instance, El-Taweel et al. estimated the energy consumption by generating a set of speed profiles using bus trip information and simulating different traffic conditions \cite{ ElTaweel2021}. Fiori et al. used electric bus data from two months of operation to develop a microscopic power-based energy consumption model \cite{Fiori2021}.

In this research, we overcome the limitations of the current models that require a large number of physical parameters or detailed and bespoke driving data on specific routes and bus powertrains by modeling the energy consumption, the emissions, and the cost with open-source transit data.
 
\subsubsection{Carbon emissions and health impacts}
\label{intro-emission-health}
Often in research, the environmental and social benefits of bus electrification have focused only on emission reductions. In a comparative life cycle assessment between diesel and electric buses in Macau, researchers found that electric buses can help reduce emissions when used with a mix of clean electricity generation, high charging efficiency, and low electricity distribution losses \cite{Song2018}. Another life cycle analysis showed that electric buses are best for reducing local air pollution and regional impact; however, the benefit depends on the environmental performance of electricity production \cite{Nordelf2019}.

Despite its importance, most techno-economic studies have not estimated the human health impacts of bus electrification. In 1970, a study showed that a 50\% reduction of air pollution in urban areas could save up to 4.5\% of economic costs related to morbidity and mortality \cite{Lave1970}. According to a study in the early 2000s, air pollution cost about 0.7\% of the Gross Domestic Product of the United States, and 94\% of that cost was attributed to the reduction in human health from fine particulate matter exposure \cite{Muller2007}. Still, in the early 2020s, Turner et al. found a significant causal link between air pollution and cancer and mortality \cite{Turner2020}. These studies indicate that we must analyze the health impact to reveal the social benefit of bus electrification. Therefore, we estimate the savings in health costs due to the reduced air pollution, as well as the reduction of greenhouse gas emissions.

\subsubsection{Total Cost of Ownership analysis}
Many studies have modeled the Total Cost of Ownership (TCO) for electric bus fleets. Khandekar et al. found that due to declining prices for lithium-ion batteries, the TCO for electric buses can be less than diesel for intra-urban routes in India, even without subsidies \cite{Khandekar2018}. Vilppo and Markkula found that it is best to invest in lithium titanate battery buses, fast charging at terminal stations, and low-power overnight charging \cite{Vilppo2015}. Another study expects the TCO for electric buses over conventional ones to decline by 2030, especially for transit agencies in California, due to changes in fuel prices, government subsidies, stringent emission standards, and enhanced performance of electric vehicles \cite{Ambrose2017}. Despite their essential insights on bus electrification, the current models in the literature are not generalizable to routes across any transit agency because of their data constraints.

\subsubsection{Analysis tools for vehicle fleet performance}
Many studies have developed simulation and software tools to support the decision-making process for bus electrification. The Future Automotive Systems Technology Simulator (FASTSim) is an advanced analysis tool that compares various vehicle powertrains, including conventional, hybrid, all-electric, and alternative fuels, on vehicle efficiency, performance, cost, and battery life \cite{Brooker2015}. The US Department of Transportation created the Transit Greenhouse Gas Emissions Estimator, a spreadsheet tool that computes life-cycle emissions from construction, operations, and maintenance activities for different transit modes \cite{FTA2016}. The US Environmental Protection Agency built another tool called the Diesel Emissions Quantifier, which estimates health benefits from the reduction in exhaust diesel emissions due to on-road and non-road sources for all the counties in the US \cite{EPA2010}. 

Lajunen et al. proposed a simulation tool that estimates energy consumption and life cycle costs from electric bus operations for different weather conditions, charging methods, and operating routes \cite{Lajunen2018}. Xu et al. developed the transit Fuel and Emissions Calculator (FEC), which compares life cycle emissions for various alternative fuels and powertrains for fleets. The calculator uses user inputs such as duty cycle, road grade, passenger loading, and location-specific energy mix and estimates the bus technology-fuel combination with the least greenhouse gas emissions \cite{Xu2015}. However, these works require domain knowledge and detailed information on the bus operation.

\subsubsection{Application of open transit data}
General Transit Feed Specification (GTFS) is a ``data specification that allows public transit agencies to publish their transit data in a format that can be consumed by a wide variety of software applications'' \cite{gtfs}. It has been widely applied to improving the planning and operation of transit systems. Many transit agencies publish their operation data in this format. 

Barbeau et al. highlighted the importance of maximizing the investments in the GTFS data for creating tools in trip planning, ride-sharing, route visualization, time-table creation, network planning, transit accessibility, and interactive voice response technologies \cite{Antrim2013}. One recent example is the PubtraVis, a dynamic interactive visualization tool based on GTFS data that describes the geographical and statistical patterns of the public transit operation and aids the communication between transit operators, city authorities, and the general public \cite{Prommaharaj2020}. Another example is the research by Yuan et al., which combined static and real-time GTFS data with field measurements to deduce the route characteristics and travel activity to estimate the energy consumption of electric trains \cite{Yuan2018}. However, our work is the first to apply the GTFS data to evaluate public bus electrification in terms of its economic, environmental, and social benefits.

\subsection{Research Contribution}



This research aims to propose and develop a tool to analyze the economic, environmental, and social impacts of public bus electrification based on open-source data across many cities worldwide, without requiring the strenuous collection of specific data sets. The contributions of this research are:
\begin{enumerate}
    \item To use physics-informed machine learning models with readily available open-source data to evaluate the total cost of ownership, the greenhouse gas emissions, and the health impacts of bus electrification, and
    \item To demonstrate the scalability of the proposed tool with a case study of bus electrification analysis for the Massachusetts Bay Transportation Authority (MBTA) in the Greater Boston area, Massachusetts, USA, and the Agenzia Mobilita Ambiente Territorio (AMAT) in Milan, Italy.
\end{enumerate}

\begin{figure}
	\centering
		\includegraphics[width=1\columnwidth]{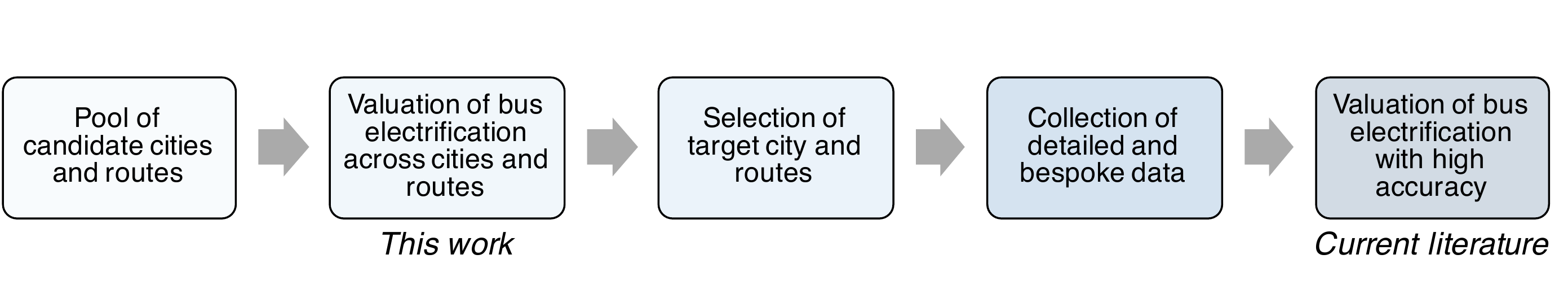}
	\caption{Research Scope}
	\label{FIG:concept}
\end{figure}

The scope of this work focuses on developing a high-level guideline to support decision-making on bus electrification; refer to Figure \ref{FIG:concept}. Our tool evaluates the value of bus electrification across cities and routes. The result can help target a specific city or a set of routes, for which detailed and bespoke data can be collected and the electrification benefits can be estimated with higher accuracy. Our method is subject to larger uncertainty in modeling accuracy than the current literature, which focuses on value estimation with granular and high-fidelity data. However, our tool allows a comparative analysis and a sensitivity analysis across many cities and routes.

\subsection{Paper Organization}
The paper is organized as follows. In Section \ref{sec:metho}, we detail the models to evaluate bus electrification in terms of energy consumption, fleet parameters, emissions, health impact, and total cost of ownership, as well as the software architecture of the proposed tool. In Section \ref{sec:casestudy}, we provide a case study of the bus electrification analysis for the Greater Boston area in Massachusetts, USA, and the Milan metropolitan area, Italy. In Section \ref{sec:discuss}, we summarize the key findings and describe the limitations of this research and future work. Appendix \ref{sec:appendix} gives a further description of the open-source data.

\section{Methodology}
\label{sec:metho}
In the following, we describe how we select and cluster the candidate routes for evaluation, which helps reduce the computation efforts of the electrification valuation. Then, we propose a physics-informed machine learning model to estimate the energy consumption of electric buses without detailed drive cycle data. We briefly describe the fuel economy model of diesel buses. We also describe models to estimate the bus fleet parameters, the emissions, the health impact, and the total cost of ownership of electrifying public transit buses. We also briefly describe the software architecture for tool development.

\subsection{Selection and Clustering of Candidate Routes}
A transit agency may have a large number of transit routes to plan and operate. For example, more than eight transit services and 150 routes are operated by MBTA and AMAT. Therefore, we define the following criteria to prioritize and select routes for electrification evaluation. First, we select routes that traverse urban areas only, which are prone to traffic congestion, because electric buses have superior performance to diesel buses particularly in stop-and-go traffic conditions \cite{Xu2015, He2018}. Second, we select routes that pass through transport hubs and Points of Interest (PoIs) that match the GTFS data. We exclude transportation modes other than the buses, such as trollies, metro trains, and trams. 

\begin{figure}
	\centering
		\includegraphics[width=0.3\columnwidth]{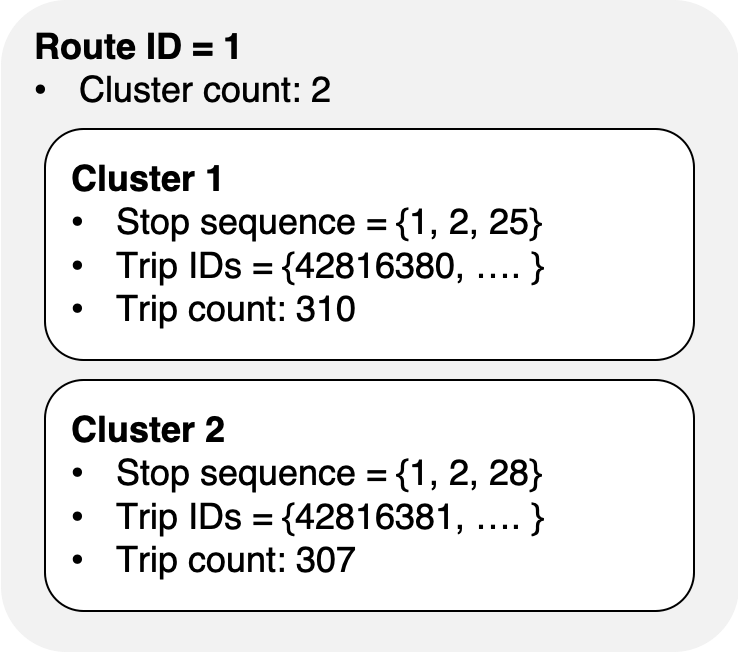}
	\caption{Example of trip clustering for a given route}
	\label{FIG:clustering}
\end{figure}

After selecting candidate routes to analyze, we cluster the trips of each route in the data set because the same route may be traversed with a different number of stops and via different paths, depending on the operation design. We define a cluster as a set of trips with a unique sequence of stops. Figure \ref{FIG:clustering} shows an example of a cluster. A route may have multiple clusters, and we calculate the number of clusters for each route and the number of trips in each cluster. Clustering reduces the computation cost, allowing us to estimate the energy consumption for each cluster instead of a large number of individual trips. In this paper, we notate $k$-th cluster of route $r$ as $\mathcal{C}_{k,r}$.

\subsection{Energy Consumption}
In the following, we describe our physics-informed machine learning model to estimate the energy consumption of electric buses. We also describe the fuel economy used to depict the energy consumption efficiency of diesel buses fro the literature. 

\subsubsection{Electric Bus}
We model the energy consumption of electric buses for a given route using only open-source data, such as GTFS, road grades, ambient temperatures, and passenger weights. We overcome the lack of time series data of speed along a route by proposing an estimation model based on a Monte Carlo simulation, a physical model of longitudinal vehicle dynamics, and a regression technique. We apply the model to the Manhattan drive cycle, an open-source time-series of speed data that depicts the driving behavior in a typical stop-and-go traffic condition of urban environments. While this drive cycle does not precisely represent all bus routes in all cities, we use this data to provide a conservative estimate for energy consumption.    

An overview of the energy consumption estimation is as follows. First, we perform a Monte Carlo simulation to sample operation scenario parameters, such as road grades, ambient temperature, and passenger weights between every two consecutive stops on a route. Second, for the sampled scenario parameters, we use a physical model of longitudinal vehicle dynamics to estimate the energy efficiency of the Manhattan drive cycle, considering it as a single trip. Third, we fit a regression model for the energy efficiency as a function of the operation scenario parameters. Fourth, we predict the energy efficiency between two consecutive stops on a route and calculate the energy consumption for a given route by aggregating the distance between two consecutive stops on the route.

In the following, we briefly explain the physical models from the literature to estimate bus energy efficiency in terms of traction, heating, ventilation, and air conditioning (HVAC), auxiliary load, and regeneration. We describe the process of generating the parameters via Monte Carlo simulation, performing a regression analysis, and estimating the total energy consumption for a given route.  

\paragraph{Physical Models of Energy Consumption}
\label{sec:metho-energy-physical}

First, we use a model by Guzzella et al. to estimate the tractive energy and describe a bus as a point mass that uses power for longitudinal movement \cite{guzzella2007vehicle}. In this model, energy is consumed for the tractive power required to move the vehicle against various opposing forces, including vehicle inertial effects, climbing grade of the road, rolling resistance at the tire, and aerodynamic drag. The tractive power from sampled point $i$ to the consecutive sampled point $j$, $P_{trac}^{i,j}$is expressed as:
\begin{align}
    & P_{trac}^{i,j} = v_i \cdot \left( F_{accel}^{i,j}  + F_{grade}^{i,j} + F_{roll}^{i,j} + F_{drag}^{i,j} \right), 
\end{align}
where $v_i$ is the velocity of the bus at point $i$, $F_{accel}^{i,j}$ is the acceleration force from $i$ to $j$, $F_{grade}^{i,j}$ is the grade climbing force from $i$ to $j$, $F_{roll}^{i,j}$ is the rolling resistance force at the tire from $i$ to $j$, and $F_{drag}^{i,j}$ is the air drag force from $i$ to $j$, and
\begin{align} 
    & F_{accel}^{i,j} = ma_i, \\
    & F_{grade}^{i,j} = mg\sin({\alpha^{i,j}}), \\
    & F_{roll}^{i,j} = mg C_{rr}\cos({\alpha^{i,j}}), \\
    & F_{drag}^{i,j} = 0.5\rho AC_d v_i^2, \\
    & a_i = \frac{v_{j} - v_{i}}{t_{j} - t_{i}}, \\
    & m = m_{bus} + m_{pass} \cdot \beta_{t}^{pass}
\end{align}
where $m$ is the total mass of the bus, $a_i$ is the acceleration at $i$, $g$ is gravitational acceleration, $\alpha^{i,j}$ is the road grade angle from $i$ to $j$, $C_{rr}$ is the coefficient of rolling resistance, $\rho$ is the air density, $A$ is the cross-sectional area of the bus, $C_d$ is the air drag coefficient, $t_i$ is the time step at $i$, $m_{bus}$ is the mass of the unloaded bus, $m_{pass}$ is the average mass of a passenger, and $\beta_t^{pass}$ is the number of passengers at time $t$. Note that the acceleration $a_i$ is estimated using the Euler central difference method from a time series of speeds, such as the Manhattan drive cycle.

When deceleration occurs, i.e., $a_i < 0$ or when the bus is moving downhill, i.e., $\sin{\alpha^{i,j}} < 0$, a negative tractive power is possible, i.e., $P_{trac}^{i,j} < 0$. This can cause regenerative braking, where the inertial kinetic and/or the gravitational potential energy of the bus can be converted and charged to the battery.

In the second step, we use the model and parameters by El-Taweel et al. \cite{ElTaweel2021} to describe the power consumption for the HVAC system at time $t$, as  $Q_t^{HVAC}$, following their variable names. Third, we use the value of the auxiliary power as $P_{aux}$ from work by \cite{Vepslinen2019}. The total non-tractive power consumed by the vehicle, $P_{non-trac}^t$, is expressed as:
\begin{align}
    P_{non-trac}^t = \frac{Q_t^{HVAC}}{\eta^{COP}} + P_{aux},
\end{align}
where $\eta^{COP}$ is the HVAC system coefficient of performance from \cite{ElTaweel2021}. 

The total energy consumption can be categorized into three cases. First is powered driving, where the tractive power is positive, and energy is discharged from the battery. Second is regenerative braking, where the sum of tractive and non-tractive power is negative and charges the battery. If their sum exceeds the battery capacities in power and energy, then charging is saturated and friction braking is used to help decelerate the vehicle. Third is idling, where no tractive power is consumed. We can reflect all conditions and describe the energy consumption as an equation. Note that in this study, we determine the points $i$'s and $j$'s so that $\Delta t$ $= t_j - t_i$ $= 0.1$ seconds, corresponding to the time interval of the Manhattan drive cycle. The energy consumption to traverse from a point $i$ to the consecutive point $j$ over a time step $\Delta t$, $\Delta E_i$ is expressed as: 

\begin{align} 
    \Delta E_i   =  \Delta t \cdot \frac{1kW}{1000W} \cdot \frac{1\text{hr}}{3600\text{sec}}  \cdot \max \left(  -P_{max,motor}, \frac{P_{trac}^{i,j}}{\eta_{battery} \eta_{motor}} + \frac{P_{non-trac}^t }{\eta_{battery} } \right), 
\end{align}
where $P_{max,motor}$ is the electric motor power capacity, $\eta_{battery}$ is the battery efficiency, and $\eta_{motor}$ is the motor efficiency.

Finally, the energy efficiency between a pair of consecutive stops, $s$, in kWh per km, $EE_s$, can be calculated. Note that there are multiple sampled points between a pair of two consecutive stops $s$, as the sampling time interval is $\Delta t=0.1$ seconds in the Manhattan drive cycle. We notate the sampled points between the pair $s$ as $I_s$ and express $EE_s$ as: 
\begin{align}
    EE_s = \frac{1000\text{m}}{1\text{km}} \frac{1}{\sum_{i\in I_s} v_i \Delta t} \sum_{i \in I_s} \Delta E_i,
    \label{eq:energy_efficiency}
\end{align} 
where $I_s$ is the set of sampled sampled in the Manhattan drive cycle.





\paragraph{Simulation of Route Energy Consumption}
Since GTFS data does not include the time series data of bus speed, we use the Manhattan drive cycle to estimate the energy consumption along a given route. We perform Monte Carlo simulation to sample three input parameters - the number of passengers, the outside temperature, and the road grade, and estimate the energy consumption of 40-feet long buses. The purpose is to address the variation in trips on different routes in different cities.

For the number of passengers, we sample from the probability distribution of ridership. Most city buses have a peak occupancy of 40 passengers. We vary the travel scenarios from an empty to a full bus, uniformly. For the outside temperature, we consider the average monthly temperatures for each city. We sample the temperature values in the Monte Carlo simulation according to a Gaussian mixture model. However, we use the average yearly temperature to evaluate the TCO. Note that we consider the lowest temperature of the monthly average to evaluate the feasibility of the electric bus in Section \ref{sec:metho_fleet_feasibility}, because the feasibility becomes more challenging with a higher HVAC energy consumption. We sample the road grade between every two consecutive stops.

We create 20,000 random samples of the three parameters, i.e., the number of passengers, outside temperature, and road grade. For these parameter samples, we simulate the Manhattan drive cycle as the cycle between each pair of consecutive stops $s$ in each trip $j$ of cluster $\overline{E_{\mathcal{C}_{k,r}}}$ and estimate the energy efficiency between the stops using Equation \ref{eq:energy_efficiency}. We fit a regression model between the three parameters and the energy efficiency values based on a six-degree elastic net polynomial function. The regression model predicts the energy efficiency over each pair of consecutive stops $s$ in trip $j$, $EE_{j, s}$. The regression statistics for the model are reported in Section \ref{sec:energy_efficiency} for the case studies on the Greater Boston and Milan areas. The energy efficiency is multiplied by the distance between the pair of stops, $d_{j,s}$. The  energy consumption of a single trip $j$, $E_j$, is expressed as:
\begin{align}
    E_j = \sum_s EE_{j,s} \cdot d_{j,s}.
\end{align}

The daily energy consumption for cluster $\mathcal{C}_{k,r}$, $\overline{E_{\mathcal{C}_{k,r}}}$, is the mean over the energy consumption of all trips $j$ in cluster $\mathcal{C}_{k,r}$, expressed as: 
\begin{align}
    \overline{E_{\mathcal{C}_{k,r}}} = \frac{1}{|\mathcal{C}_{k,r}|} \sum_{j \in \mathcal{C}_{k,r}} E_j.
\end{align} 
where $|\mathcal{C}_{k,r}|$ is the cardinality of  set $\mathcal{C}_{k,r}$, i.e. the number of clusters of route $r$.

The annual energy consumption of route $r$, $E_r^\text{annual}$, is expressed as:
\begin{align}
    E_r^\text{annual} = \sum_{\mathcal{C}_{k,r}} \overline{E_{\mathcal{C}_{k,r}}} \cdot \frac{\text{365 days}}{\text{1 year}}.
\end{align}

Note that the daily energy consumption varies between weekdays and weekends because the number of buses in operation and the number of passengers (and consequently their weight on the bus) vary. However, the daily and annual averages will suffice to analyze the emissions, the health impact, and the total cost of ownership.

\subsection{Diesel Bus}
In this paper, we use the fuel economy of diesel buses and use it to estimate the emissions, the health impact, and the total cost of ownership, instead of developing a physical model for the energy consumption of diesel buses. The fuel economy $FE$ is defined in miles per gallon (MPG) based on the average speed on the route, $\overline{v_r}$ using the following polynomial regression model:
\begin{align}
    & FE = -0.0032 [c_v \cdot \overline{v_r}]^2 + 0.2143 [c_v \cdot \overline{v_r}] + 0.9726, \label{eq:fuel_economy}
\end{align}
where $c_v$ is the conversion factor from km/hr to miles/hr. This equation is obtained from chassis dynamometer data for various real-world driving cycles that included the impact of heating/cooling loads and terrain on fuel economy \cite{clark2009assessment}. Note that $\overline{v_r}$ is to be derived in Equation \ref{eq:speed_2}.

\subsection{Bus Fleet Parameters}
Although GTFS data provides limited information on the operation, we can extract useful parameters of fleet operation to estimate the electrification benefits. These parameters include the total number of electric buses needed, the average speed on a route, the total vehicle miles traveled, the estimated road grade between stops, the number of chargers needed, and the range feasibility.

\subsubsection{Number of buses}
To estimate the capital and operational cost of electric buses, we estimate the maximum number of buses required to serve the current transit schedule. We must estimate the bus fleet size because GTFS does not include information on the bus fleet, nor does each transit agency openly publish the bus allocation data. For each bus that stops at each stop on a route, we estimate the cycle length (or duration of trip) as the difference between the arrival time of the bus at the last stop on the route and the departure time of the bus at the first stop on the route. We also estimate the headway of the bus at a stop as the difference between the arrival times of the subject bus to the previous bus.

However, the cycle length and headway values vary with time and stop on the route. Therefore, we take a conservative approach and estimate the number of buses for each route as follows: 
\begin{align}
    & N_{r, in}(s,t) = \frac{C_{in}(s,t)}{h_{in}(s,t)}  , \\
    & N_{r, out}(s,t) = \frac{ C_{out}(s,t) }{ h_{out}(s,t)},\\
    & N_{r, in}^v = \max_s (\max_t ( N_{r, in}(s,t) )), \\
    & N_{r, out}^v = \max_s (\max_t ( N_{r, out}(s,t) )), 
\end{align}
where $N_{r, in}(s,t)$ and $N_{r, out}(s,t)$ are the estimated numbers of buses at time $t$ and stop $s$ on the route $r$ for in-bound and out-bound directions, respectively, $C_{in}(s,t)$ and $C_{out}(s,t)$ are the cycle lengths, in time, estimated for the bus that arrives at stop $s$ at time $t$ for in-bound and out-bound directions, respectively, $h_{in}(s,t)$ and $h_{out}(s,t)$ are the headways, in time, estimated for the bus that arrives at stop $s$ at time $t$ for in-bound and out-bound directions, respectively, and $N_{r, in}^v$ and $N_{r, out}^v$ are the maximum numbers of buses on route $r$ for in-bound and out-bound directions, respectively. Finally, the number of buses required for route $r$, $N_r^v$ is expressed as:
\begin{align}
    N_r^v = N_{r, in}^v + N_{r, out}^v.
\end{align}

In other words, we estimate the number of buses at each time, each stop of the route, and each direction of the route. For each direction, we take the maximum number of buses over all times and all stops, and then sum them to produce the total number of buses required for the route.  

\subsubsection{Average speed of a route}
We estimate the average speed for the $k$-th cluster of route $r$, $\overline{v_{\mathcal{C}_{k,r}}}$, as: 
\begin{align}
    \overline{v_{\mathcal{C}_{k,r}}} = \frac{ \sum_{j\in \mathcal{C}_{k,r}} \sum_s d_{j,s} }{  \overline{C_{\mathcal{C}_{k,r}}}} \cdot \frac{60 \text{min}}{1 \text{hr}}, \label{eq:speed_1}
\end{align}
where $d_{j,s}$ is the distance between a pair of stops $s$ in trip $j$ in cluster $\mathcal{C}_{k,r}$ and  $\overline{C_{\mathcal{C}_{k,r}}}$ is the sample mean of the cycle lengths for trips in the $k$-th cluster of route $r$. The average speed of a route $r$, $\overline{v_r}$, is defined as the average of the cluster speeds, i.e., 
\begin{align}
    \overline{v_{r}} = \frac{1}{|\mathcal{C}_{k,r}|} \sum_{k} [\overline{v_{\mathcal{C}_{k,r}}}].
    \label{eq:speed_2}
\end{align}

\subsubsection{Vehicle kilometers traveled}
The vehicle kilometers traveled on a given route $r$ annually, $\text{VKT}_r^\text{annual}$ is calculated as:
\begin{align}
    \text{VKT}_r^\text{annual} = \sum_{k} \sum_{j\in \mathcal{C}_{k,r}} \sum_s d_{j,s},
\end{align}
where the trip samples $j$ belong to a given year. Note that the sum over index $k$ indicates the sum over all clusters $k$ of route $r$. The sum over $k$ index is the sum over all clusters $k$ of route $r$. Also added a notation for $\mathcal{C}_{k,r}$ in Nomenclature, which means $k$-th cluster of route $r$.

\subsubsection{Road grade between stops}
The road grade between two points, $\alpha^{i,j}$ can be estimated as:
\begin{align}
    \alpha^{i,j} = \arcsin{ \left( \frac{\Delta e_{i,j}}{d_{i,j}} \right)} 
    , \label{eq:grade_1}
\end{align}
where $\Delta e_{i,j}$ is the change in elevation and $d_{i,j}$ is the distance between two points, $i$ and $j$. We obtain the elevation and distance data from the Google API (refer to Appendix \ref{sec:appendix}). Note that sometimes trips skip consecutive stops, although GTFS does not describe the skipped stops. We assume that when consecutive stops have the same values for arrival time, the first stop is visited, but the rest of the stops are skipped.

\subsubsection{Number of chargers}
To estimate the number of chargers required we assume that the electric buses charge at the depot overnight. We omit the detailed modeling of the charging process, such as the charging dynamics, the state of charge of buses at arrival and departure, and relocating buses in and out of charging plugs. The number of chargers required for route $r$, $N^c_r$, is estimated as:
\begin{align}
    N^c_r = \sum_{k} \frac{\overline{E_{\mathcal{C}_{k,r}}}}{P_c \cdot T_c \cdot \eta_c} \frac{1000 \text{W}}{1 \text{kW}} , \label{eq:charger}
\end{align}
where $\overline{E_{\mathcal{C}_{k,r}}}$ is the daily total energy consumption for cluster $\mathcal{C}_{k,r}$,  $P_c$ is the charging power, $T_c$ is the fastest charging time, and $\eta_c$ is the charging efficiency. We assume that the charging power $P_c$ is the minimum of the bus charging power capacity and the charger charging power capacity. We consider charging efficiency as a user input to the tool, which determines the number of chargers as an output. 


\subsubsection{Range feasibility}
\label{sec:metho_fleet_feasibility}
It is possible that a route is infeasible to electrify for a given battery capacity. We assume that a route is feasible if the following is condition is satisfied:
\begin{align}
    {\overline{E_{\mathcal{C}_{k,r}}}} < E_{max,b} \cdot N_{\mathcal{C}_{k,r}}, \label{eq:feasibility}
\end{align}
where $E_{max,b}$ is the energy capacity of the battery. We assume all routes are feasible for diesel buses. Namely, diesel buses have sufficient range to cover all routes before refueling.

\subsection{Emissions and Health Impact}
Electric buses can significantly reduce emissions thus alleviating the climate impacts of transportation and improve human health. In this paper, we focus on two types of emissions: (i) carbon dioxide ($\text{CO}_2$) emissions to estimate the climate impact of city bus operation and (ii) particulate matter (PM2.5) emissions to estimate the human health impacts \cite{Muller2007}. The following describes the modeling approach for $\text{CO}_2$ emissions and the cost of public health posed by the on-road PM2.5.

\subsubsection{$\text{CO}_2$ emissions} 
We calculate the well-to-wheel emissions for electric and diesel buses. Note that the equation for the fuel economy is described in Equation \ref{eq:fuel_economy}. The equation for the well-to-wheel $\text{CO}_2$ emissions of a diesel bus, for a given route, $\text{CO}_2^{d,r}$, is calculated based on the emission factors in metric tons per year:
\begin{align}
\begin{split} 
    \text{CO}_2^{d,r} =  \text{VKT}^\text{annual}_r \cdot \left[  EF^{CO_2, d}_{W2T} \cdot \frac{1000kg}{1g} +   EF^{CO_2, d}_{T2W} \cdot  \frac{1}{FE} \cdot c_v  \right],
\end{split}
\end{align}
where $EF^{CO_2, d}_{W2T}$ is the well-to-tank emission factor for a diesel bus and $EF^{CO_2, d}_{T2W}$ is the tank-to-wheel emission factor for a diesel bus.

The well-to-wheel $\text{CO}_2$ emissions for an electric bus on a given route $r$, $\text{CO}_2^{e,r}$, is expressed in metric tons per year:
\begin{align}
    \text{CO}_2^{e,r} = \sum_{\mathcal{C}_{k,r}} \overline{E_{\mathcal{C}_{k,r}}} \cdot EF^{CO_2, e}_{W2T}\cdot \frac{365 \text{days}}{\text{year}},
\end{align}
where $EF^{CO_2, e}_{W2T}$ is the well-to-tank emission factor for an electric bus.
Note that electric buses have zero tailpipe CO$_2$ emissions, i.e., zero tank-to-wheel emissions.

\subsubsection{Health impact from PM2.5}
PM2.5 emissions from conventional fuel use are a root cause of chronic respiratory diseases caused by poor air quality as discussed in Section \ref{intro-emission-health}. We evaluate the benefit of bus electrification by monetizing the health impact of PM2.5 emissions. The PM2.5 exhaust for diesel buses for a route $r$, $\text{PM}^{d,r}$ is expressed in grams per year as:
\begin{align}
    \text{PM}^{d,r} =  \text{VKT}^\text{annual}_r \cdot EF^{PM2.5, d}_{T2W},
\end{align}
where $EF^{PM2.5, d}_{T2W}$ is the tank-to-wheel emission factor of PM2.5 for a diesel bus. The PM2.5 exhaust for electric buses is zero. We omit the non-exhaust PM2.5 emissions from the tire, brakes, and road materials  Although some studies argue they contribute significant portions of PM2.5 emissions \cite{Beddows2021, Grigoratos2014}, those emissions may be similar between electric and diesel buses and therefore omitted in our model. We also do not consider PM2.5 emissions from well-to-tank, assuming that those emissions from processing diesel or generating electricity are far from the cities of interest. 

From the PM2.5 emissions for each route, we calculate the US dollar value corresponding to the loss of life from tailpipe emissions from diesel buses. We estimate the health impact of the PM2.5 emissions of a diesel bus for a given route $r$, $HI_r$, as:
\begin{align}
    & HI_r = PI \cdot FF \cdot VSL, \\
    & PI = IF \cdot \text{PM}^{d,r} \cdot \frac{\text{kg}}{1000\text{g}}, 
\end{align}
where $PI$ is the population intake in kg of PM2.5 inhaled by humans, $FF$ is the effect factor in disability-adjusted life years lost per kg of PM2.5 inhaled, $VSL$ is the value of statistical life in million US dollars, and $IF$ is the intake fraction, i.e., the fraction of emissions inhaled by an exposed population from a given source. The research in \cite{Fantke2019, Apte2012} report the impact of PM2.5 emissions across 3,448 cities around the globe and has identified city and country-specific values of $FF$ and $IF$ for use.

\subsection{Total Cost of Ownership}
We model the economic benefit of electrifying transit buses in terms of the TCO, which can be compared to the TCO of diesel buses for a given route. We aim to identify which routes are most economical and best suited for electrification.

\subsubsection{Capital cost}
\paragraph{Electric bus:}
The capital cost of bus electrification for route $r$, $\text{CAPEX}_{e,r}$, considers the buses and the charging infrastructure as:
\begin{align}
    \text{CAPEX}_{e,r} = c_{e,bus} \cdot N_{r}^v +  (c_{c,install} + c_{c,unit}) \cdot N_{r}^c.
\end{align}
The charger installation costs may depend on the site readiness and electrical equipment. In the following case study of Section \ref{sec:casestudy}, we include the average cost of labor, material, permit, and taxes required for installing from 6 to 50kW chargers per site.

\paragraph{Diesel bus:}
We assume the refueling infrastructure for diesel buses already exists and calculate the capital cost of only diesel vehicles for route $r$, $\text{CAPEX}_{d,r}$ as:
\begin{align}
    \text{CAPEX}_{d,r} = c_{d,bus} \cdot N_{r}^v.
\end{align}

\subsubsection{Operation and maintenance costs}
Operations and maintenance costs may vary with transit agency size. Therefore, we make assumptions common to public transit. Typical bus manufacturers offer 12 years of bus and battery warranty. Therefore, we omit the mid-life overhaul costs for equipment such as battery, motor and inverter replacement, or engine replacement for diesel buses. Many electric bus manufacturers use lithium iron phosphate batteries, which have a longer life and sustain a larger number of charging cycles. If buses charge every day across a 12-year lifetime, this will result in approximately 4,380 cycles. Modern lithium iron phosphate cells have a cycle life ranging from 5,000 to 12,000 cycles depending on the operating conditions, covering 12 years.  

\paragraph{Electric bus:} 
Electric buses typically have lower operation and maintenance costs due to cheaper charging and maintenance. The charging cost paid to the utility may include two parts: the energy cost and the demand charge. The demand charge penalizes the peak power consumption at a facility, often calculated over a month. The operation cost of electric buses has three components - the annual energy cost of route $r$ for year $y$, $C_{kWh,r}^y$, the annual demand charge of route $r$ for year $y$, $C_{DC,r}^y$, and the annual operation and maintenance cost of electric buses of route $r$ for year $y$, $C_{O\&M,e,r}^y$. They are expressed as follows:

\begin{align}
    & C_{kWh,r}^y = c_{kWh} \cdot (1+r_{kWh})^y \cdot \frac{E_r^{annual}}{\eta_c}\\
    & C_{DC,r}^y = c_{DC}\cdot N_r^c \cdot P_c  \cdot (1+r_{DC})^y \cdot \frac{\text{12 months}}{\text{year}}\\
    &C_{O\&M,e,r}^y = c_{O\&M,e}^v \cdot N_r^v \cdot c_v  \cdot \text{VKT}_r^{annual}+ c_{O\&M}^c \cdot N_r^c,
\end{align}
where $c_{kWh}$ is the electricity purchase cost per unit energy, $E_r^\text{annual}$ is the annual energy consumption of route $r$, $r_{kWh}$ is the percentage change in unit energy price each year, $c_{DC}$ is the demand charge per power charged each month, $r_{DC}$ is the percentage change in demand charge each year, $c_{O\&M,e}^v$ is the operation and maintenance cost of an electric bus per mile, $\text{VKT}_r^\text{annual}$ is the annual vehicle kilometers traveled for route $r$, and $c_{O\&M}^c$ is the operation and maintenance cost of a charger. Note that we assume the peak power occurs when all chargers are used simultaneously at full power, $P_c$. Note, the demand charges can be reduced by incorporating on site storage \cite{Woo2021}, but we do not consider on-site storage in this study.

\paragraph{Diesel bus:}
The operation cost of diesel buses for year $y$ has two components - the annual fuel cost of route $r$ for year $y$, $C_\text{fuel,r}^y$, and the annual operation and maintenance cost of diesel buses for route $r$ for year $y$,  $C_{O\&M,d,r}^y$. They are expressed as the following:

\begin{align}
    & C_\text{fuel,r}^y = c_{fuel}  \cdot (1+r_f)^y \cdot c_v \cdot \frac{\text{VKT}_r^\text{annual}}{FE}, \\
    & C_{O\&M,d,r}^y = c_{O\&M,d}^v \cdot N_r^v \cdot c_v\cdot \text{VKT}_r^{annual} , 
\end{align}
where $c_{fuel}$ is the fuel price per mile, $FE$ is the fuel economy, $r_f$ is the percentage change in fuel price each year, and $c_{O\&M,d}^v$ is the operation and maintenance cost of a diesel bus per mile.


\subsubsection{Salvage cost}
Electric buses, diesel buses, and chargers can provide some returns from reuse and recycling. Though Neubauer et al. find potential in creating revenue with the second use of an electric battery, some challenges lie ahead \cite{Neubauer2012}. We do not include second life battery resale in the current study but only express the salvage costs for electric and diesel buses, $C_{e,salv,r}$ and $C_{d,salv,r}$ respectively, as: 
\begin{align}
    & C_{e,salv,r} = -[r_v \cdot c_{e, bus} \cdot N_r^v + r_{c} \cdot c_{c,unit} \cdot N_r^c ] ,\\
    & C_{d,salv,r} = - [r_v \cdot c_{d, bus} \cdot N_r^v],
\end{align}
where $r_v$ is the residual value of a bus in percentage, and $r_c$ is the residual value of a charger in percentage.

\subsubsection{Total Cost of Ownership}
The TCO is calculated as the capital, operation, and maintenance costs for each year. In this study we assume 12 years of ownership. The net present value (NPV) of TCO for route $r$ for electric bus fleet, $\text{TCO}_{e,r}^\text{NPV}$ is:
\begin{align}
    \begin{split}
        \text{TCO}_{e,r}^\text{NPV} = & \text{CAPEX}_{e,r} + \frac{(1+r_d)^{12} - 1}{r_d\cdot(1+r_d)^{12}} \cdot C_{O\&M,e,r}^y   + \sum_{y=1}^{12} \left[ C_{kWh,r}^y + C_{DC,r}^y \right]   + \frac{1}{(1+r_d)^{12}} C_{e,salv,r}.
    \end{split}
    \label{eq:tco_npv_electric}
\end{align}
The net present value (NPV) of TCO for route $r$ for diesel bus fleet, $\text{TCO}_{d,r}^\text{NPV}$ is:
\begin{align}
    \begin{split}
        \text{TCO}_{d,r}^\text{NPV} = & \text{CAPEX}_{d,r}  + \frac{(1+r_d)^{12} - 1}{r_d\cdot(1+r_d)^{12}} \cdot C_{O\&M,d,r}^y   + \sum_{y=1}^{12}  C_{fuel,r}^y   + \frac{1}{(1+r_d)^{12}} C_{d,salv,r},
    \end{split}
    \label{eq:tco_npv_diesel}
\end{align}
where $r_d$ is the discount rate.  


\subsection{Software Development}

We implement the modeling and analysis described above into an on-demand TCO calculator for fleet electrification, available at \url{https://www.ocf.berkeley.edu/~electrify/}. The online tool takes user inputs and provides the value of fleet electrification in terms of the cost, the emissions, the technical feasibility, and the health impacts per route. The following describes the input, the output, and the software architecture of the online tool.

The online tool has three user inputs:  
\begin{itemize}
    \item City and Routes: The user can select the routes to analyze for the chosen city
    \item Cost parameters: The energy purchase cost $c_{kWh}$, the demand charge $c_{DC}$, the diesel price $c_{fuel}$, the electric bus cost $c_{e,bus}$, and the percentage change in demand charge each year for the transit agency $r_{DC}$
    \item Fleet parameters: The length of the bus in feet and the charging efficiency $\eta_c$
\end{itemize} 

The online tool has the following outputs:
\begin{itemize}
    \item TCO, $\text{CO}_2$ emissions, and the monetized health impact of deploying diesel buses on the selected routes
    \item TCO, $\text{CO}_2$ emissions, and the monetized health impact of deploying electric buses on the selected routes 
    \item If on-route or fast charging is required, beyond overnight depot charging
    \item Technical feasibility of the bus electrification
\end{itemize}

\begin{figure*}
	\centering
		\includegraphics[width=0.6\textwidth]{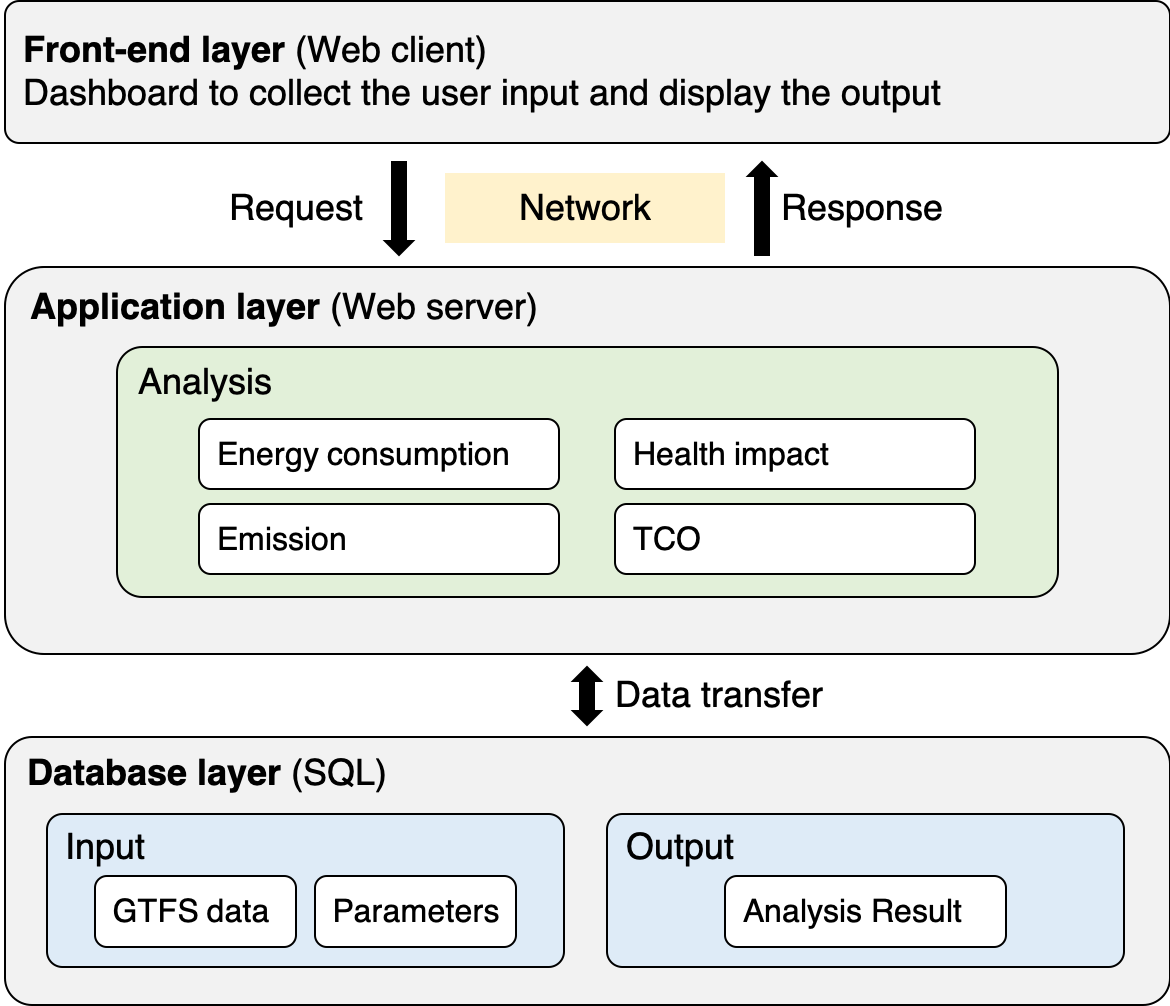}
	\caption{Software Architecture for the Online Tool}
	\label{FIG:web_server}
\end{figure*}

The online tool has an architecture with three main components - front-end, application, and database, as shown in Figure \ref{FIG:web_server}. In the front layer, we build a dashboard that collects the user input and displays the results. In the application layer in Python, we calculate in real-time the energy consumption, the emissions, the health impact cost, and the TCO. In the database, we store the data, such as the user-input data, GTFS data on routes, trips, and stops, and the output of the analysis.

\section{Case Study: Boston and Milan}
\label{sec:casestudy}

We present a case study with the proposed tool to estimate the value of bus electrification for the transit agencies in the Greater Boston area, Massachusetts, USA, and the Milan metropolitan area, Italy. In the following, we briefly describe the scope of the case study. For each route of the transit agencies, we estimate the energy efficiency, the TCO, the health impact, and the greenhouse gas emissions. We analyze the results among the routes for each city and between the two cities. While the tool can analyze all bus sizes, we consider 40-feet long buses because they are the most common size.

\subsection{Cities of Study}

\begin{figure*}
	\centering
		\includegraphics[width=1\textwidth]{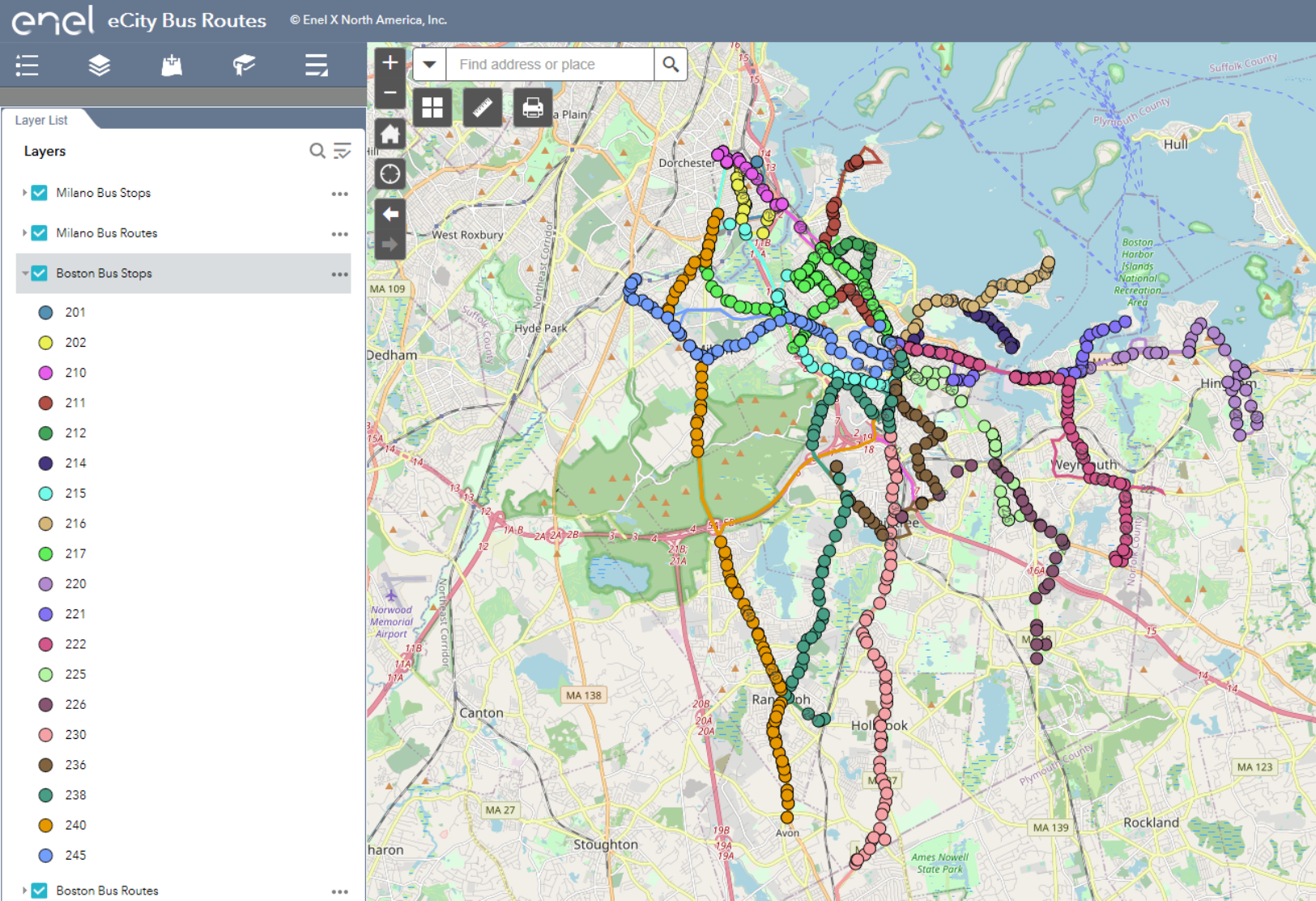}
	\caption{Route display for the Massachusetts Bay Transportation Authority (MBTA) case}
	\label{FIG:routedisplay_boston}
\end{figure*}

\begin{figure*}
	\centering
		\includegraphics[width=1\textwidth]{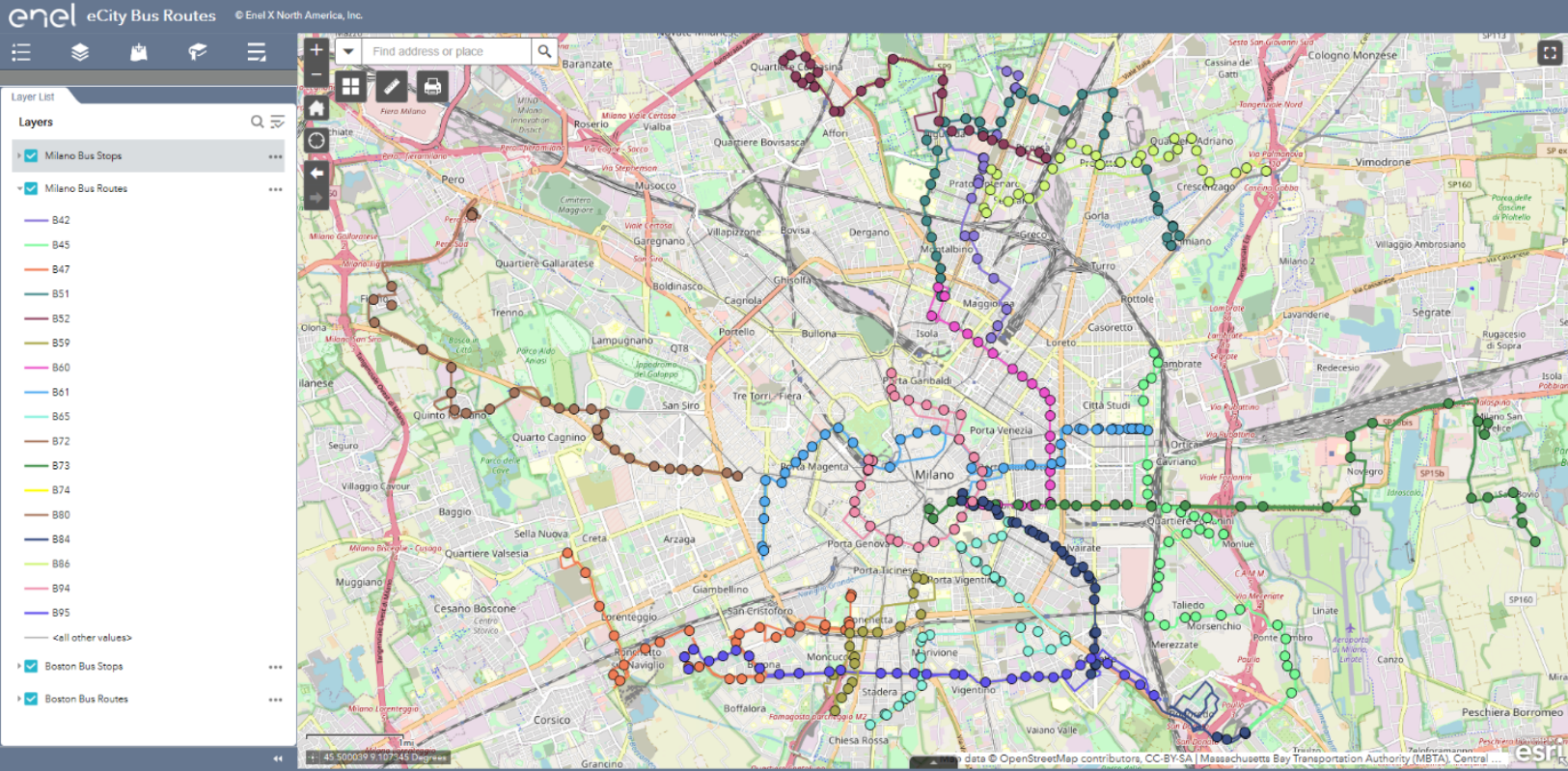}
	\caption{Route display for the Agenzia Mobilita Ambiente Territorio (AMAT) case}
	\label{FIG:routedisplay_milan}
\end{figure*}

We evaluate the value of bus electrification for two public transit agencies - Massachusetts Bay Transportation Authority (MBTA) and Agenzia Mobilita Ambiente Territorio (AMAT). The MBTA services the Greater Boston area and operates 171 bus routes and four rapid transit routes \cite{mbta_intro}. The average ridership on a weekday for buses was about 387,000 before the COVID 19 pandemic, according to the data in January 2020 \cite{mbta_dashboard}. The AMAT services the metropolitan area of Milan and publishes the GTFS data under a contract with the Municipality of Milan (Comune di Milano).

\begin{figure*}
	\centering
		\includegraphics[width=1\textwidth]{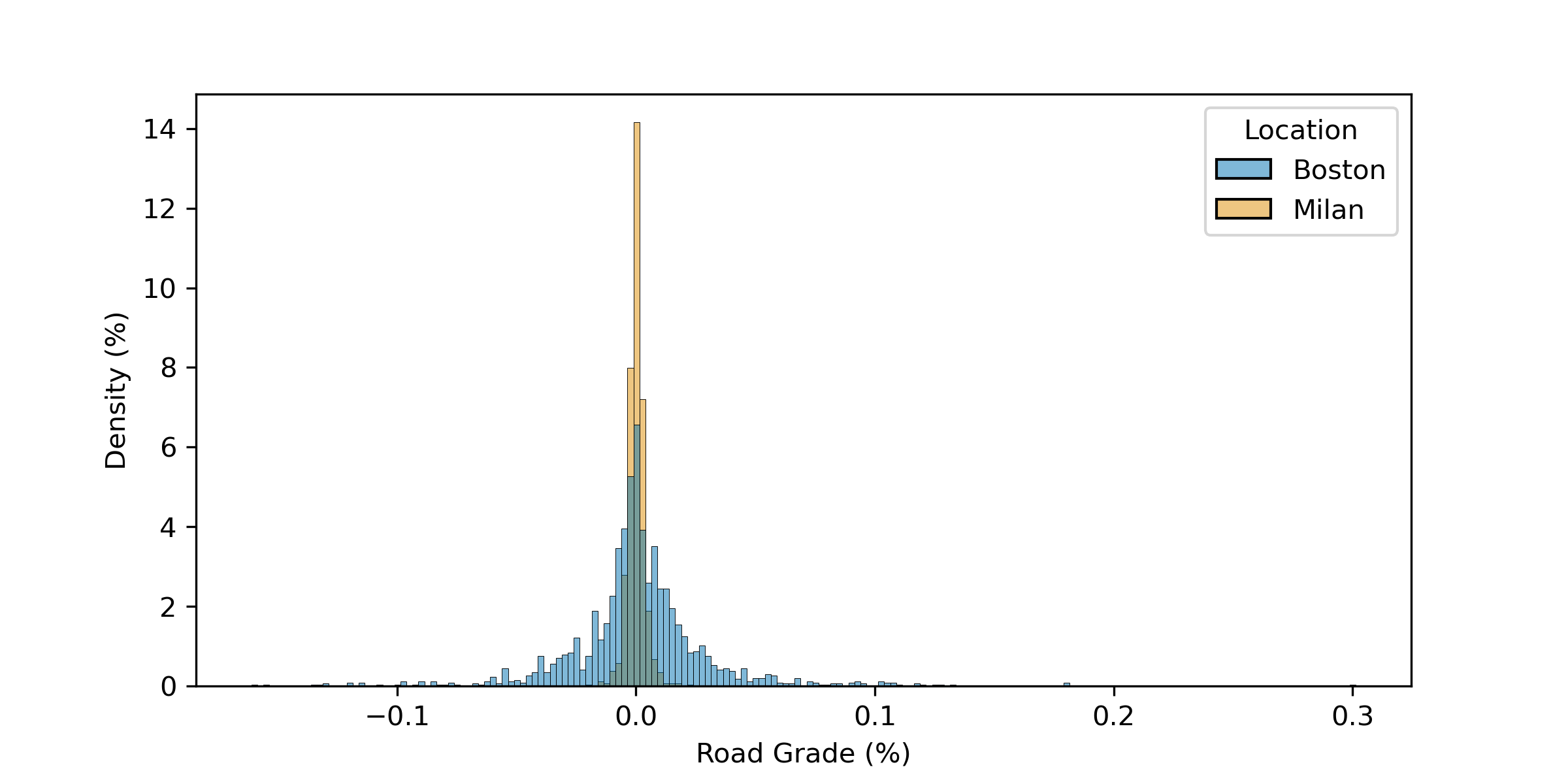}
	\caption{Road Grade Distribution for Bus Routes of MBTA (Boston) and AMAT (Milan)}
	\label{FIG:grade_dist}
\end{figure*} 

The GTFS data was collected from February 27th to March 14th in 2020 for the MBTA, and from March 14th to March 25th in 2020 for the AMAT. The routes chosen for MBTA and AMAT are visualized in Figure \ref{FIG:routedisplay_boston} and Figure \ref{FIG:routedisplay_milan}, respectively. For MBTA, we use the Quincy garage routes pertaining to line numbers 201, 202, 210, 211, 212, 214, 215, 216, 217, 220, 221, 222, 225, 226, 230, 236, 238, 240, and 245. For AMAT, we use the routes of line numbers B42, B45, B47, B51, B52, B59, B60, B61, B65, B73, B80, B84, B86, B94, and B95. The total number of buses to operate the chosen routes is estimated as 93 for MBTA and 89 for Milan. The average values of road grade in the Greater Boston and Milan areas are both 0.000\%. However, the standard deviation of the Greater Boston area is 0.029\%, which is larger than that of Milan with 0.003\%. The minimum and maximum values of the road grade in the Greater Boston area are -0.161\% and 0.301\%, respectively. The minimum and maximum values in the Milan area are -0.020\% and 0.022\%, respectively. The distribution of the road grade values is presented in Figure \ref{FIG:grade_dist}.

\subsection{Energy Efficiency}
\label{sec:energy_efficiency}
We estimate the energy efficiency with Equation \ref{eq:energy_efficiency} in Section \ref{sec:metho-energy-physical} to compare routes of different lengths. {The regression model takes three variables that can contribute significantly to the energy efficiency of the bus, namely the ambient temperature, the passenger load, and the road grade. As described in the following, the high accuracy of the regression result in the following shows that its feature engineering is successful.} 

For the regression to estimate energy efficiency, we use a six-degree elastic net polynomial function. This  model is trained on a Monte Carlo simulation of a physics-based energy model, using open data from MBTA and AMAT routes. The total sample size is 20,000 simulated drive cycles, each for MBTA and AMAT. The samples are split into 80\% and 20\% sets for the training and test sets, respectively. 
For Milan, the predicted efficiency values ranged from -0.349 to 2.185 kWh/km for different sets of input parameters, i.e., the number of passengers, the outside temperature, and the road grade. The root mean square error (RMSE) is 0.00085 kWh/km for the test set, which is in the order of 0.1\%. For Boston, the predicted efficiency values ranged from -8.777 to 10.169 kWh/km. The RMSE is 0.00204 kWh/km for the test set, which is in the order of 0.01\%. The small RMSE values indicate that the physics-informed machine learning model provides high-accuracy estimates of energy efficiency, relative to a physics-based model.


\begin{figure*}
	\centering
		\includegraphics[width=.8\textwidth]{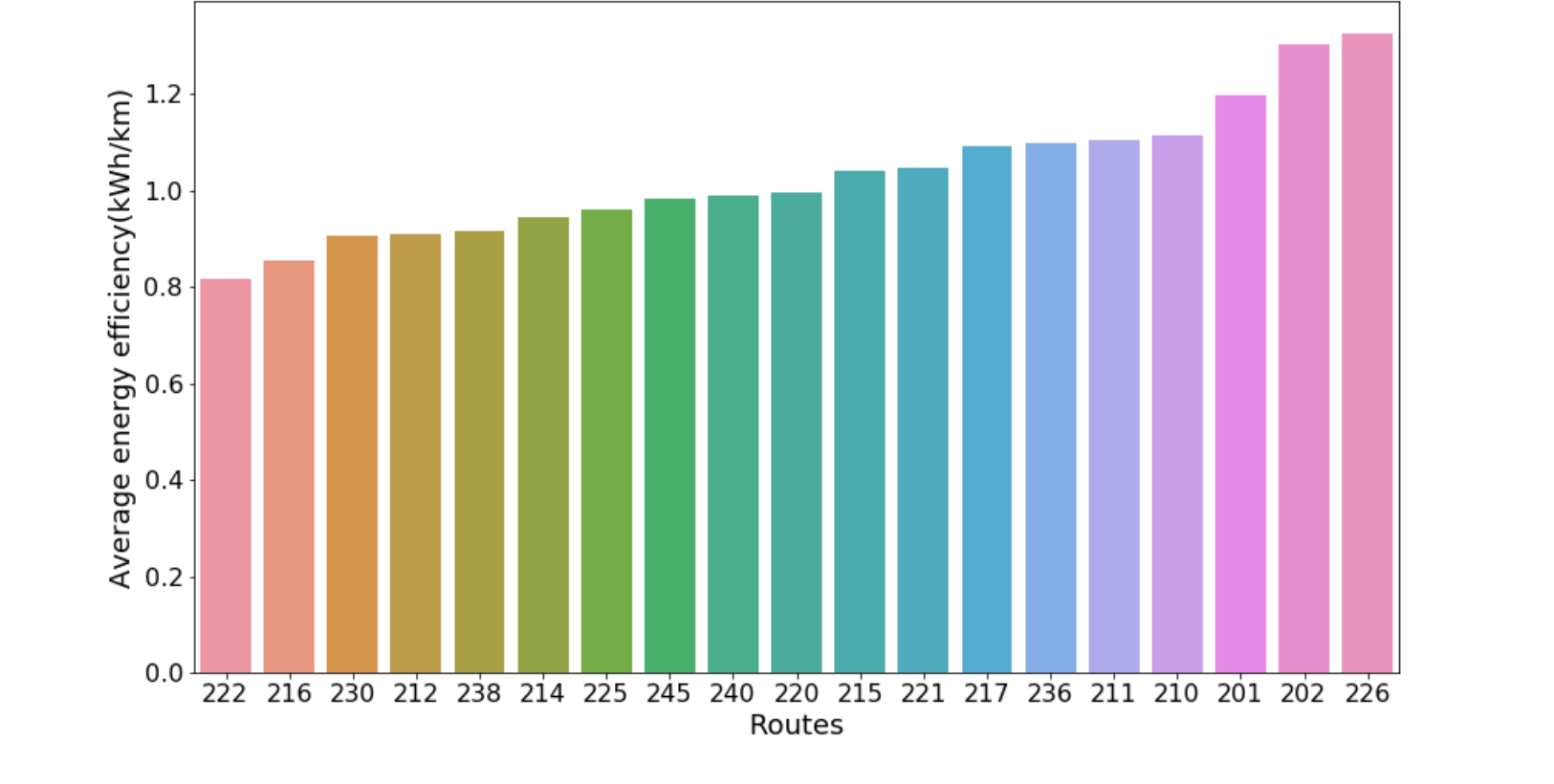}
	\caption{Energy efficiency of 40-foot long electric buses for MBTA}
	\label{FIG:boston_energyefficiency}
\end{figure*} 

\begin{figure*}
	\centering
		\includegraphics[width=.8\textwidth]{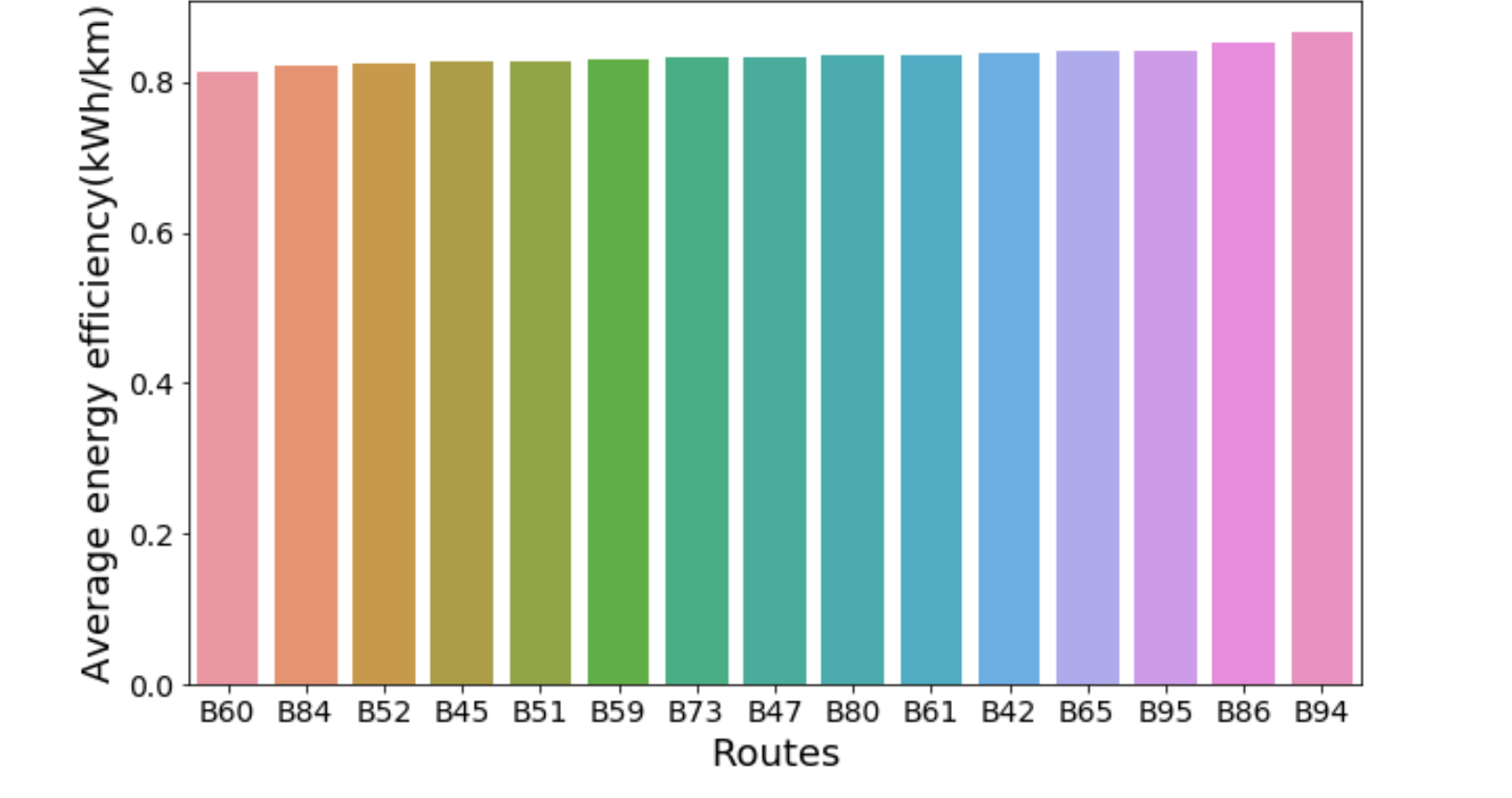}
	\caption{Energy efficiency of 40-foot long electric buses for AMAT}
	\label{FIG:milan_energyefficiency}
\end{figure*}

Based on the regression result, the energy efficiency of electric buses for each route is estimated and presented for MBTA and AMAT in Figure \ref{FIG:boston_energyefficiency} and Figure \ref{FIG:milan_energyefficiency}, respectively, in the increasing order of kWh/km. Route 222 in MBTA is the most energy efficient route at 0.81 kWh/km. Routes 206 and 202 are estimated to be the least efficient at 1.32 and 1.30 kWh/km, respectively. The variation in energy efficiency is more significant for MBTA compared to AMAT. Most routes for AMAT show similar energy efficiency values, ranging between 0.82 to 0.88 kWh/km. This is possibly because the road grade in the Greater Boston area has a larger variability than that of Milan. Note that the energy efficiency is estimated only for electric buses in this study, though the total cost of ownership (including the fuel cost), the health impact, and the emissions are assessed for both diesel and electric buses and described next.



\subsection{Total Cost of Ownership}

\begin{sidewaysfigure*} 
    \centering 
        \includegraphics[width=.75\textwidth]{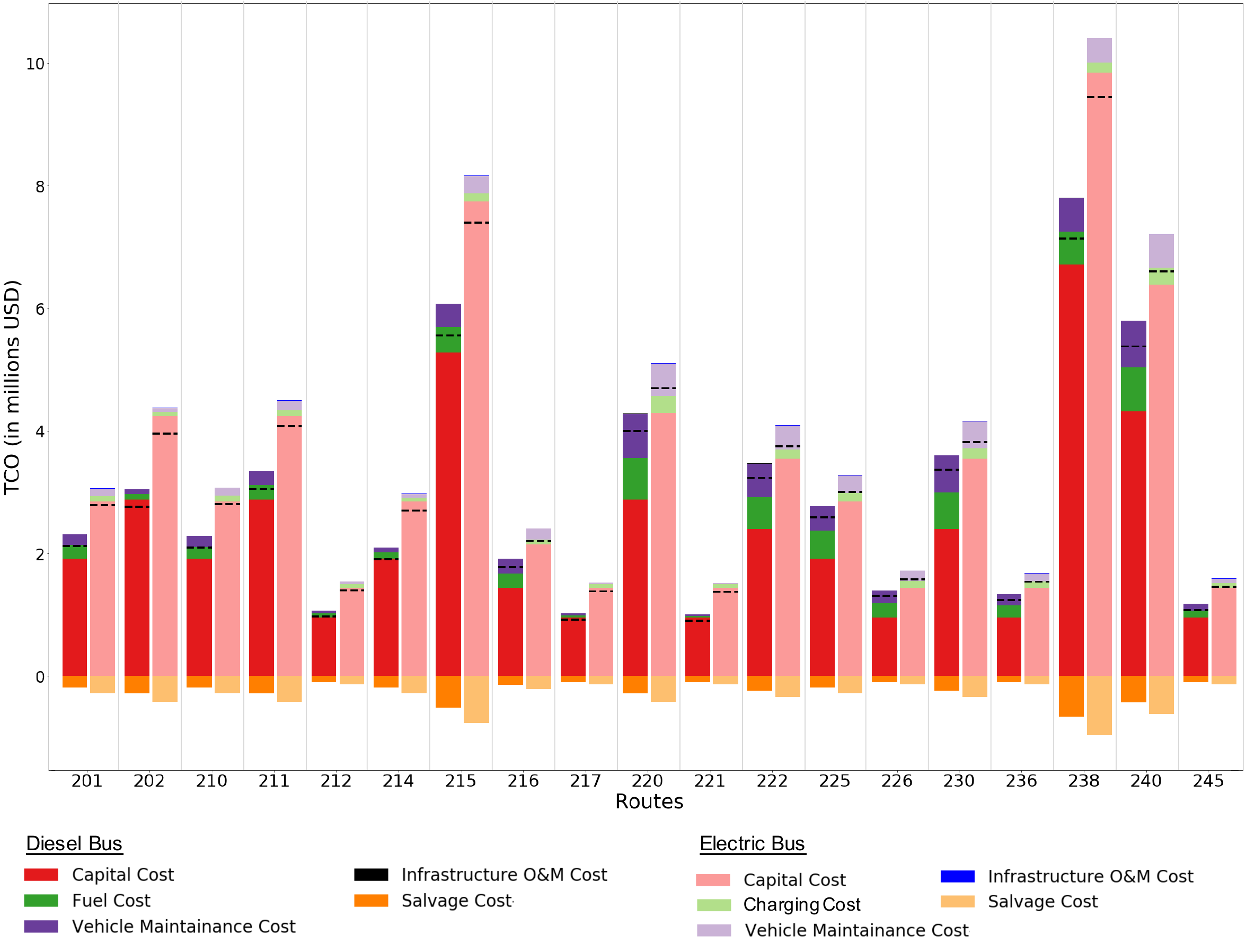} 
    \caption{Net Present Value of Total Cost of Ownership for MBTA}
    \label{FIG:TCO_boston}
\end{sidewaysfigure*}

\begin{sidewaysfigure*}
    \centering 
        \includegraphics[width=.7\textwidth]{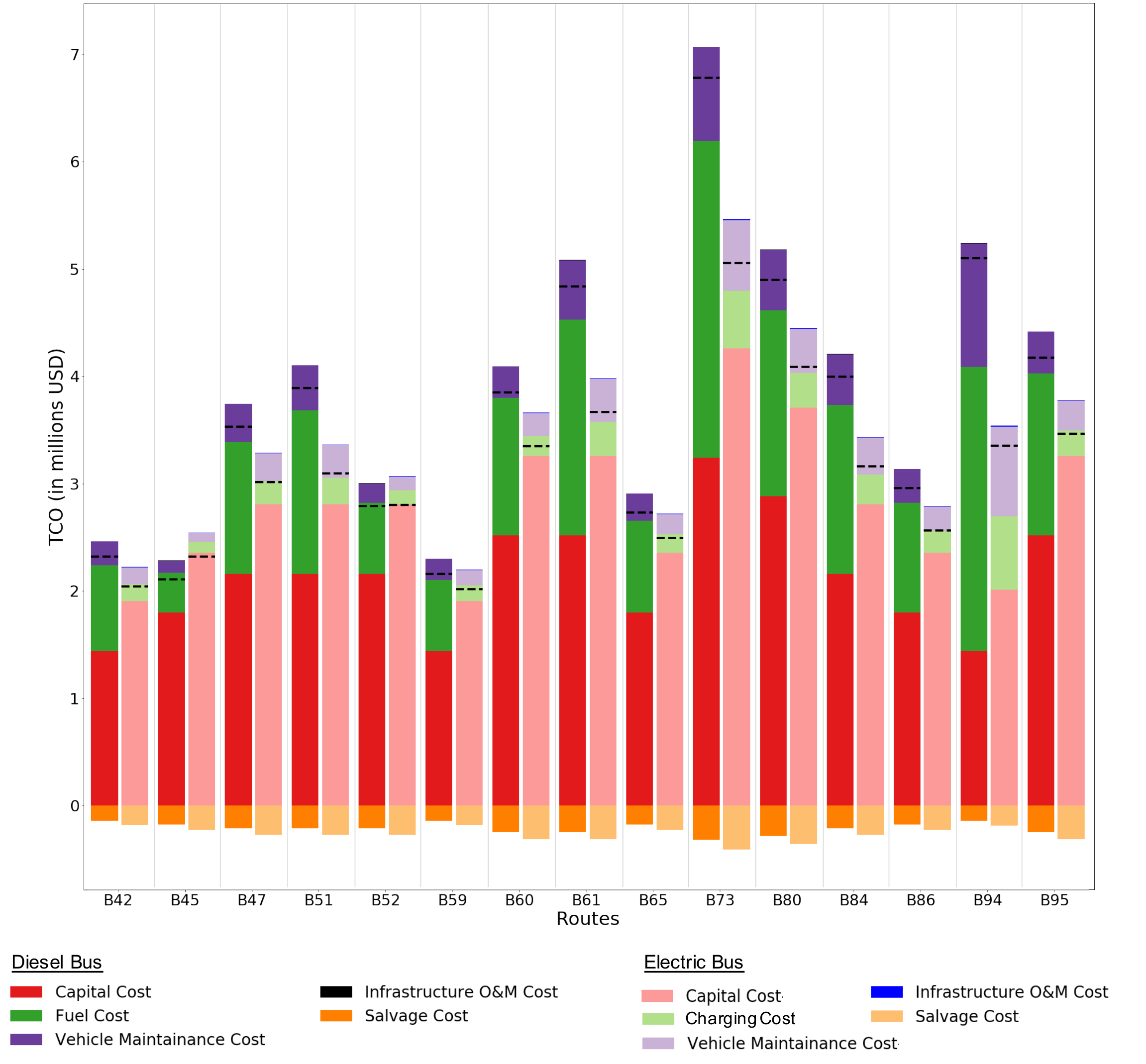} 
    \caption{Net Present Value of Total Cost of Ownership for AMAT}
    \label{FIG:TCO_milan}
\end{sidewaysfigure*}


To estimate TCO, we use the parameter values in Table \ref{tab:tco_param}, although they are user inputs that can be changed in the online tool. The NPV of TCO for various routes in MBTA and AMAT are shown in Figure \ref{FIG:TCO_boston} and Figure \ref{FIG:TCO_milan}, respectively, using Equations \ref{eq:tco_npv_electric} and \ref{eq:tco_npv_diesel}. The individual components of the TCO are color-coded, as shown in the legend. We show the diesel bus TCO on the left bars and the electric bus TCO on the right bars for each route, for easy comparison. The net TCO is annotated by dashed black lines, accounting for the costs and the revenue (or salvage cost).

\begin{table}[t]
\caption{Parameters Used for the TCO Estimation (User Input to the Tool)}
\label{tab:tco_param}
\centering 
\begin{tabular}{l l l }
\toprule
Parameter & MBTA & AMAT \\
\midrule
Energy price, $c_{kWh}$, (USD/kWh) & 0.098 & 0.232 \\
\midrule
Demand charge, $c_{DC}$, (USD/kW) & 8 & 8 \\
\midrule
Fuel Price, $c_{fuel}$, (USD/gallon) & 2.546 & 5.8 \\
\midrule
\makecell[l]{Electric bus capital cost, $c_{e,bus}$, (USD)} & 750,000 & 450,000 \\ 
\midrule
\makecell[l]{Rate of increase in demand charge $r_{DC}$ (\%)} & 0 & 0 \\
\midrule
Charging efficiency, $\eta_c$ (\%) & 95 & 95 \\
\midrule
Bus size & 40 feet & 40 feet \\
\bottomrule
\end{tabular}
\end{table}

For the inputs used in this analysis, all studied routes in the Greater Boston area have a larger TCO for electric buses than for diesel buses. This is primarily due to the higher capital costs of electric buses, which is the largest contributor to the TCO. Note that in this case study, we omit the impact of government subsidies on bus fleet electrification \cite{FTA2021}, which could significantly reduce the TCO - an important policy implication for these results. A secondary factor is the relatively low price of diesel fuel in Boston, which reduces the operating costs of diesel buses.  We argue that routes with a larger ratio of operation cost to the capital cost have a better economic potential for bus electrification, since the operation cost savings could provide an earlier payback for the capital cost. 


In the case of Milan, all routes except B45 and B52 have smaller TCO values with electric buses than with diesel buses. This contrasts with the Great Boston area. The capital costs for purchasing electric buses are 450,000 USD in Milan, much smaller than 750,000 USD in the Greater Boston area. Moreover, the cost of diesel is 5.8 USD per gallon in Milan compared to 2.5 USD per gal in the Greater Boston area. Therefore, we learn that bus electrification becomes appealing when the electric bus capital cost is low, and the diesel fuel cost is high.   

\subsection{Health Impacts}

\begin{figure*}
	\centering
		\includegraphics[width=.9\textwidth]{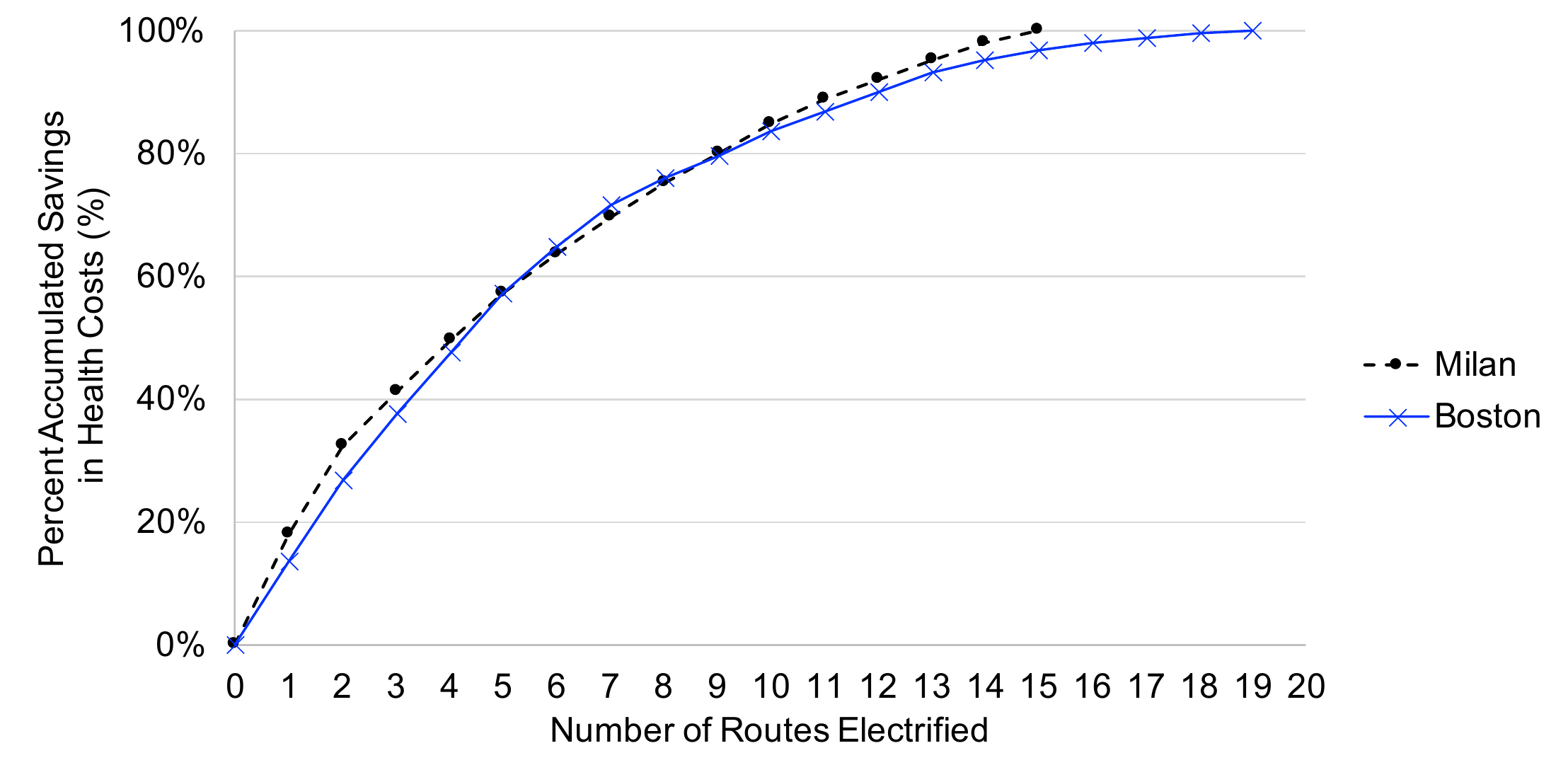}
	\caption{Percent savings of health costs by bus electrification}
	\label{FIG:healthimpact}
\end{figure*}

We demonstrate the use of our tool to support decision-making on bus electrification based on the health impacts. We monetize the health cost of bus electrification as a function of the total distance traveled. We present the health benefits, estimated as the percent savings in health costs as a function of the number of routes electrified, in Figure \ref{FIG:healthimpact}. We display the routes in order of decreasing health benefits and calculate the cumulative effect of electrifying each route. For MBTA, about 90\% of savings in health costs can be achieved by electrifying the first 13 routes out of 19. For AMAT, electrifying the first 11 routes out of 15 can save about 90\% of the total health costs. In fact, almost half of the bus emission-related healthcare costs can be eliminated by electrifying just 4 routes in both cities.

\subsection{Cost vs. Emissions}

\begin{figure*}
	\centering
		\includegraphics[width=.9\textwidth]{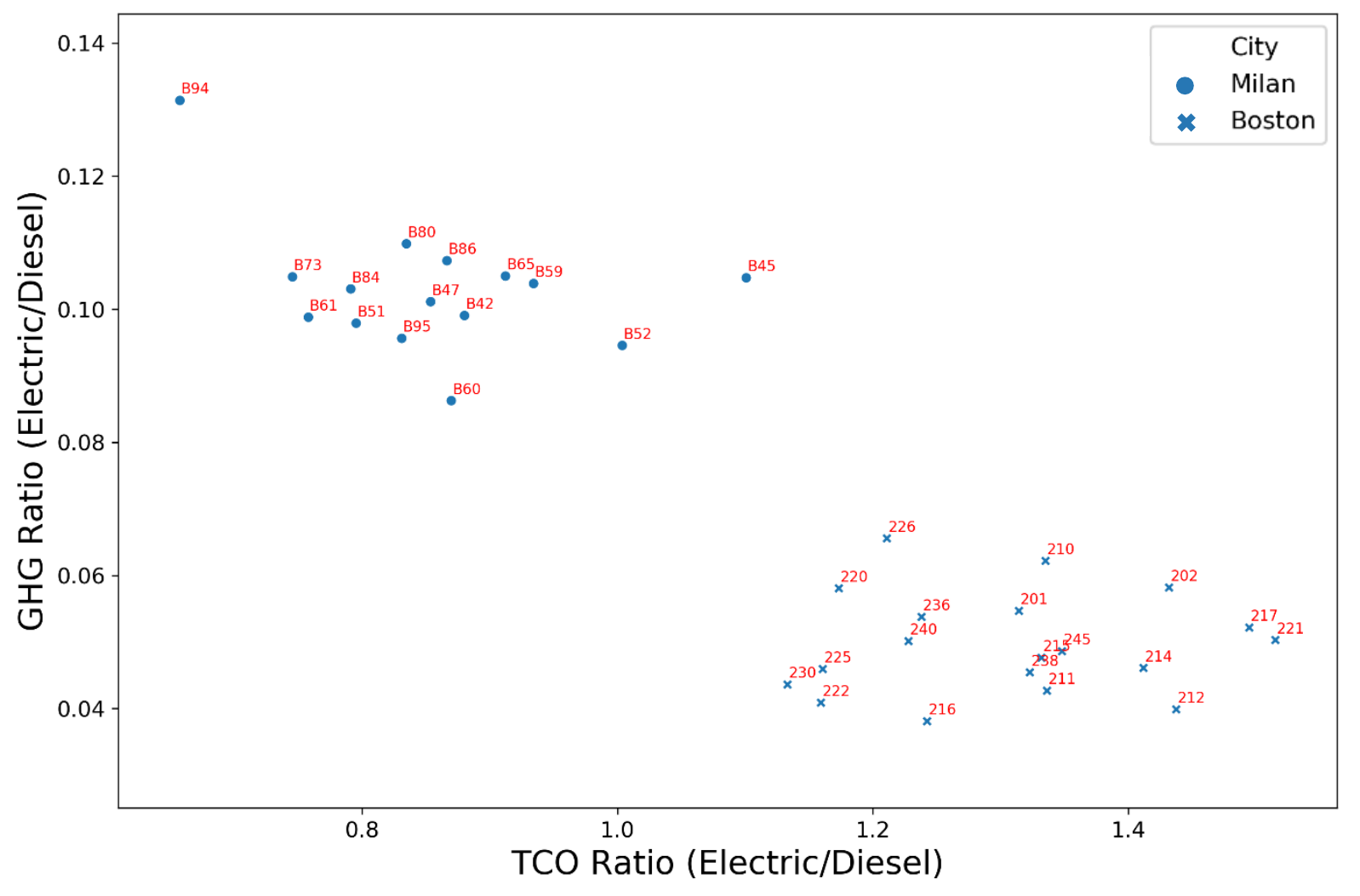}
	\caption{Total Cost of Ownership and Greenhouse Gas emissions as a ratio of electric buses to diesel buses}
	\label{FIG:tco_vs_emission}
\end{figure*}

We exemplify another use of our tool to identify which routes offer the highest potential with electrification when comparing the performance with multiple objectives. We evaluate the environmental and economic benefits, in terms of the greenhouse gas emissions and the TCO. In Figure \ref{FIG:tco_vs_emission}, the x-axis shows the TCO ratio from electric buses to diesel buses, and the y-axis shows the greenhouse gas emissions from electric buses to diesel buses. For both axes, values less than one indicate better performance for electric buses relative to diesel. Each point represents the data of each route, annotated with the route number and a marker symbol for the city. With smaller values along the y-axis, MBTA shows a larger benefit in emissions reductions relative to AMAT. With smaller values along the x-axis, AMAT shows a larger benefit in TCO reductions relative to MBTA. 

For MBTA, routes 230, 222, and 216 are top candidates for electrification as they are on the Pareto frontier. Namely, no other routes can yield greater reductions in both TCO and GHG emissions if they are electrified. In other words, electrifying these routes is not strictly dominated by electrifying any other routes. On these routes, electric buses will produce less than 5\% of the diesel bus emissions and will cost less than 125\% of the diesel bus TCO. 

For AMAT, routes B60, B95, B51, B61, B73, and B94 are top candidates for electrification as they are also on the Pareto frontier. Most AMAT routes yield a smaller TCO with electric buses than with diesel buses, while reducing the emissions to less than 14\% of the diesel bus emissions. Note from Figure \ref{FIG:milan_energyefficiency} that route B94 has the least energy efficiency, and from Figure \ref{FIG:TCO_milan} route B94 has the highest ratio of operation cost to capital cost. This results in the least greenhouse gas emissions reductions but also the lowest TCO with electrification.

\section{Discussion}
\label{sec:discuss}
In this work, we develop a novel tool to estimate the economic, environmental, and social values of electrifying public transit buses. The tool is unique as it only requires open-source data, such as General Transit Feed Specification (GTFS), and some user inputs. Therefore, the tool overcomes the limited scalability of state-of-art models that require granular and specific data for each bus route of each transit agency, such as the drive cycle, the heating and cooling energy, and the road grade on the route. The proposed approach thus enables planners to assess bus electrification for multiple cities across the world. The contributions of this work are a) the novel modeling of public bus electrification in terms of the Total Cost of Ownership (TCO), the greenhouse gas emissions, and the health impacts, based on open-source data, and 2) the demonstration of the proposed tool with a case study for the Massachusetts Bay Transportation Authority (MBTA) in the Greater Boston area, Massachusetts, USA, and the Agenzia Mobilita Ambiente Territorio (AMAT) in Milan, Italy. The modeling innovations include a physics-informed machine learning model of the electric bus energy efficiency, clustering of trips for bus routes, and incorporating open maps and weather data. 

The case study demonstrates the scalability of the proposed tool. The results show that due to the larger variance in the road grade in the Boston area than the Milan metropolitan area, MBTA shows a larger variance in energy efficiency among its routes than AMAT. For MBTA, the TCO increases with bus electrification due to the higher capital cost of electric buses and the lower cost of diesel. For AMAT, the TCO decreases with bus electrification for most routes. For both MBTA and AMAT, a significant reduction in PM2.5 brings substantial savings in health costs. Namely, electrifying just four routes in both cities cuts in half the bus emissions-related health care costs. By analyzing the Pareto frontier of the TCO and GHG emissions, we also identify which routes offer the highest potential for electrification. The tool is publicly available to help disseminate the model: \url{https://www.ocf.berkeley.edu/~electrify/}.

In this study, we make assumptions to estimate the TCO and emissions based on open-source data. The limitations and opportunities for future improvement are as follows. First, the current model does not estimate the number of electric buses based on energy consumption measurements on the routes. Instead, we estimate the number of buses from the current operation on the busiest day at the busiest stop, thus ensuring the fleet size covers the worst case operation. Future studies might consider dispatching buses to different routes, depending on daily ridership demand and/or depending on estimated energy consumption.

Second, we assume the number of chargers is proportional to total daily energy demand for each route, thus providing a conservative estimate. This ignores the number of buses in each route and their corresponding individual daily energy consumption. The required number of chargers may be much smaller in practice as an agency would optimize the operation of bus scheduling and allocation. Therefore, this study may have overestimated the capital cost of bus electrification. Nevertheless, the model helps to screen which routes potentially yield the greatest benefits from electrification. Then a subsequent more detailed analysis can be performed on those routes, considering more sophisticated route dispatching and charge scheduling strategies.

Third, we acknowledge the limitation of GTFS data and the use of the Manhattan drive cycle. More granular drive cycle data may enable higher accuracy in estimating the energy consumption estimation, the on-route charging operation, and battery degradation. Once again, we emphasize that the proposed open data approach enables scalable route screening, after which bespoke drive cycle data can be obtained for the targeted bus routes.

Fourth, we do not consider subsidies or incentives for bus electrification in the cost analysis. Each city may have its specific carbon credit policies. The tool's estimates can be improved by incorporating relevant policy implications on the capital and operation costs and the electricity pricing. 

Fifth, the TCO and GHG estimates can improve by considering additional factors, such as variable renewable energy penetration in the power grid, fluctuation in fuel prices, cheaper maintenance costs of electric buses, the re-usability of second-life batteries, declining battery and electric bus prices, and optimized charging infrastructure planning and operation.

{Sixth, the scope of this paper does not include the life-cycle analysis of electric buses or detailed modeling of the electricity generation to charge the electric buses. Instead, we use conversion factors for well-to-tank and tank-to-wheel emissions. It is found that with an optimal operation, electric fleet operation can greatly improve the economic and environmental performance of electricity generation \cite{woo2021b}. The current work can improve by incorporating the sources and methods of electricity generation for the overall economic, environmental, and social costs.}

\section*{Acknowledgement}
This research was funded by Enel Foundation.

\appendix
\section{Open-source Data}
\label{sec:appendix}

We provide a brief description of the open-source data used to develop the proposed tool and the case study for Boston and Milan. We describe General Transit Feed Specification and the contextual data, such as the number of passengers, road grade, bus specification, emission factor, intake fraction, effect factor, the value of statistical life, and weather.

\subsection{General Transit Feed Specification (GTFS)}
GTFS is a standard data formation that transit agencies worldwide use to share their operation data. Some mandatory files for the data include agency information, stops, routes, trips, and stop times. The open format is protected under a Creative Commons license and freely available from many transit agencies.  

\subsection{Number of passengers }
The number of riders for each route is not available in GTFS, and we collect it from the open data provided by the transit agencies. For the case study of the Greater Boston area, the blue book open data portal provides the data of riders on-boarding and unloading for each stop and each trip along a given route. We obtain this data similar to the case of Milan. We calculate the average number of passengers by aggregating the riders for all trips on a route.

\subsection{Road grade and distance between stops}
We use Google APIs to estimate the road grade and the distance between stops. We obtain the elevations at each stop from the Google Elevation API to estimate the road grade. For the distance, we use the Google Distance Matrix API. We input the coordinates of two consecutive stops, for which the path and the distance are returned. The total distance traveled for a given trip is the sum of distances between two consecutive stops on the trip, assuming that the bus travels on the path from the Google API through all stops without taking a shortcut. This approach is more accurate than using Euclidean distances between the stop coordinates. 

\subsection{Bus specification }
There are various sizes of electric buses to consider, such as 30-foot, 40-foot, and 60-foot buses. The proposed model does not explicitly require the bus size, as long as the necessary data for the bus model is provided, including the mass of the bus, the energy capacity of the battery, cross-sectional area, electric motor power capacity, and fastest charging time.

\subsection{Emission factors}
 
We collect the values of emission factors from the current literature to estimate the emissions of diesel and electric buses. We omit the non-exhaust PM2.5 emissions from the tire and road materials, as they may be similar between electric and diesel buses and therefore omitted in our model.

\subsection{Intake Fraction}
 
Intake fraction is the fraction of emissions inhaled by an exposed population from a given source. This value depends on various factors, including the size of the exposed population, the distance between the source of emission and the impacted population, and the environmental persistence. We use values from the literature to estimate the exposure concentrations from ground-level emission sources, such as vehicles.

\subsection{Effect Factor}
Effect factor relates the population exposure to the associated health effects, expressed in years of life lost (YLL) or disability-adjusted life years (DALY) per kilogram intake. We use the global marginal effect factor estimates from \cite{Fantke2019}, which quantifies the change in mortality to the changes in the mass of PM 2.5 inhaled in different locations.  

\subsection{Value of Statistical Life}

Value of Statistical Life (VSL) is the monetary value that people are willing to pay for small reductions in their mortality risks \cite{epa_mortality}. This value varies across different countries, depending on cultural norms, lifestyle patterns, income levels, consumer behavior, and wage risk sensitivity. We use the Value of Statistical Life values for the USA and Italy in 1995 US dollars \cite{miller2000variations} by converting them to the present value of 2020 for the case study. The paper presents the VSL values for 13 countries across the globe.

\subsection{Weather} 
We find the yearly average and the lowest yearly temperature from open sources, such as \cite{us_climate}, and estimate the bus energy consumption that varies in different weather due to the heating and cooling demands of HVAC. For Boston, the yearly average and the yearly lowest temperatures are 11.0 and -5 degrees Celsius, respectively. For Milan, the yearly average and the yearly lowest temperatures are 14.5 and 0 degrees Celsius, respectively.

\bibliographystyle{IEEEtran} 
\bibliography{BIBLIOGRAPHY}
\end{document}